\documentclass[onecolumn,showpacs,preprintnumbers]{revtex4}
\usepackage{mathrsfs}
\usepackage{graphicx}
\usepackage{subfigure}
\usepackage{dcolumn}
\usepackage{bm}
\usepackage{epsfig}
\usepackage{amsmath,amssymb}
\usepackage{array}
\usepackage{tabularx}

\begin{document}
\title{\textbf{SEARCH FOR THE CHARMED BARYONIUM AND DIBARYON STRUCTURES VIA THE QCD SUM RULES}}
\author{Xiu-Wu Wang$^{1,2}$}
\email{wangxiuwu2020@163.com}
\author{Zhi-Gang Wang$^{1}$}
\email{zgwang@aliyun.com}

\affiliation{$^1$ School of Mathematics and Physics, North China
Electric Power University, Baoding, 071003, People's Republic of
China\\$^2$ School of Nuclear Science and Engineering, North China Electric Power University, Beijing, 102206, People's Republic of China}

\begin{abstract}
In the present work, we construct eight six-quark currents to study the $\Sigma_c\bar{\Sigma}_c$ baryonium and $\Sigma_c\Sigma_c$ dibaryon states via the QCD sum rules. For either $\Sigma_c\bar{\Sigma}_c$ baryonium or $\Sigma_c\Sigma_c$ dibaryon states, we construct four currents with the $J^P=1^-, 1^+, 0^+$ and $0^-$. Except the current $1^+$ of $\Sigma_c\Sigma_c$, we find the Borel windows for the other seven. There are two possible $\Sigma_c\bar{\Sigma}_c$ molecular states or compact six-quark states with the $J^P=1^-$ and $0^-$. The results of the current $0^+$ of $\Sigma_c\Sigma_c$ show a possible $\Sigma_c\Sigma_c$ molecular state with $20$ MeV binding energy. We find two $\Sigma_c\bar{\Sigma}_c$ baryonium resonances states with the $J^P=1^+, 0^+$ and two $\Sigma_c\Sigma_c$ dibaryon resonances states with the $J^P=0^-, 1^-$, respectively. Pole residues of the seven states are also calculated.
\end{abstract}

\maketitle

\begin{large}
\noindent\textbf{1 Introduction}
\end{large}

The dibaryon is composed of two baryons and the baryonium is formed by a baryon and anti-baryon. In 1980s, the research of the interaction between the two baryons caused great attention both theoretically and experimentally \cite{Strakovsky}. The first known dibaryon state discovered in 1932 has been the deuteron, the bound state of proton and neutron \cite{Harold}. Except for the deuteron, another genuine dibaryon resonance observed in experiment is the $d^*(2380)$ \cite{Bashkanov, Adlarson1, Adlarson2}. The baryon consists of three quarks, six quarks for the dibaryon or baryonium, it is natural to extend the concept of hexaquarks for which the quarks sitting in a common quark bag. The data of the electromagnetic form factors of the state are possible to differentiate whether it is a molecular type or a compact six-quark entity. For the deuteron, its charge radius is 2.1 fm, the proton and the neutron inside the deuteron are on average 4 fm apart from each other and do not overlap \cite{vires,Hclement}, these data depict a clear physical picture of the hadronic molecule for the deuteron. One could also make the differentiation via the decay branching, in Ref. \cite{NN1}, the decay width of $d^*(2380)$ is considered via the chiral quark model, it is proposed that $d^*(2380)$ spends $\frac{2}{3}$ of its time as a six-quark and the rest as a $\Delta\Delta$
molecule. Theoretically and experimentally, the ways which may find the difference are there, however, seldom work has been done yet for both the $\Sigma_c\Sigma_c$ dibaryon and $\Sigma_c\bar{\Sigma}_c$ baryonium in this aspect. At present, it is too hard for us to differentiate $\Sigma_c\Sigma_c$ dibaryon or $\Sigma_c\bar{\Sigma}_c$ baryonium from the compact six-quark state. We now could only say: it may be the molecular state or compact six-quark state if the mass is below the threshold of the two baryon constituents (bound state), as for the dibaryon resonance state, its mass is above the threshold of the two baryon constituents.

The oldest prediction of the existence of dibaryon with nearly $100$ MeV binding energy based on $\rm{SU}(6)$ symmetry was done by Dyson and Xuong \cite{DysonXuong} which is in good agreement with the experimental observations of $d^*(2380)$ \cite{Bashkanov, Adlarson1, Adlarson2}. Although the insufficient experimental data for the solid observations of the dibaryon and baryonium states, lots of theoretical calculations have been done in search of them. In recent years, several QCD-inspired models are applied to study the dibaryons and baryoniums,
such as the quark delocalization color screening model \cite{Xia,Huang1}, solving the Bethe-Salpeter equation
via effective Lagrangians \cite{Lu,Zhu1}, constituent quark model \cite{Huang2,Carames}, lattice QCD \cite{Morita1,Morita2}, the QCD sum rules \cite{Kodama,Chen,Wang2,Wan} and so on. Especially, in Refs. \cite{DongGuo1,DongGuo2}, the authors study the charmed hadrons to investigate the mutual interactions between them and search for poles by solving the Bethe-Salpeter equation. It is natural to ask: what are the predictions of the interactions of these charmed hadrons in the framework of the QCD sum rules? In Ref. \cite{WZG}, the scalar and
axialvector $\Xi_{cc}\Sigma_c$ dibaryon states are studied in consideration of two-baryon scattering states.
It is proposed that the two-baryon scattering states cannot saturate the QCD sum rules and their contributions can be neglected. We follow this idea and do not consider the two-baryon scattering states. The earlier study of the hidden-charm tetraquark molecular states has previously been published in a preprint \cite{WZGNNN}. In this article, we construct eight six-quark local interpolating currents to study both the $\Sigma_c$$\bar{\Sigma}_c$ baryoniums with the $J^P=1^-, 1^+, 0^+$, $0^-$ and the $\Sigma_c$$\Sigma_c$ dibaryons with the $J^P=0^+, 0^-, 1^+$, $1^-$, respectively.

The article is arranged as follows: we derive the QCD sum rules of the dibaryon and baryonium states in Sect.2; we present the numerical results and discussions in Sect.3; Sect.4 is reserved for our conclusions.

\begin{large}
\noindent\textbf{2 The QCD sum rules for the baryonium and dibaryon states}
\end{large}

Now, let us write down the corresponding two-point correlation functions from the eight constructed currents,
\begin{eqnarray}
\notag &&\Pi_{1;\alpha\beta}(p)=i\int d^4x e^{ip\cdot x}\langle 0 | J_{1,\alpha} (x) J_{1,\beta}^{\dag}(0) | 0\rangle \, ,\\
\notag &&\Pi_{2;\alpha\beta}(p)=i\int d^4x e^{ip\cdot x}\langle 0 | J_{2,\alpha}(x)J_{2,\beta}^{\dag}(0) | 0\rangle \, ,\\
\notag &&\Pi_{3}(p)=i\int d^4x e^{ip\cdot x}\langle 0 | J_3 (x) J_3^{\dag}(0) | 0\rangle \, ,\\
\notag &&\Pi_{4}(p)=i\int d^4x e^{ip\cdot x}\langle 0 | J_4 (x) J_4^{\dag}(0) | 0\rangle \, ,\\
\notag &&\Pi_{5}(p)=i\int d^4x e^{ip\cdot x}\langle 0 | J_5 (x) J_5^{\dag}(0) | 0\rangle \, ,\\
\notag &&\Pi_{6}(p)=i\int d^4x e^{ip\cdot x}\langle 0 | J_6 (x) J_6^{\dag}(0) | 0\rangle \, ,\\
\notag &&\Pi_{7;\alpha\beta}(p)=i\int d^4x e^{ip\cdot x}\langle 0 | J_{7,\alpha} (x) J_{7,\beta}^{\dag}(0) | 0\rangle \, ,\\
&&\Pi_{8;\alpha\beta}(p)=i\int d^4x e^{ip\cdot x}\langle 0 | J_{8,\alpha}(x)J_{8,\beta}^{\dag}(0) | 0\rangle \, ,
\end{eqnarray}
where
\begin{eqnarray}
\notag && J_{1,\alpha}=\bar{J}_{\Sigma}\gamma_{\alpha}J_{\Sigma} \, ,\\
\notag && J_{2,\alpha}=\bar{J}_{\Sigma}\gamma_5\gamma_{\alpha}J_{\Sigma} \, ,\\
\notag && J_{3}=\bar{J}_{\Sigma}J_{\Sigma} \, ,\\
\notag && J_{4}=\bar{J}_{\Sigma}\gamma_5J_{\Sigma} \, ,\\
\notag && J_{5}=J_{\Sigma}^TC\gamma_5J_{\Sigma} \, ,\\
\notag && J_{6}=J_{\Sigma}^TCJ_{\Sigma} \, ,\\
\notag && J_{7,\alpha}=J_{\Sigma}^TC\gamma_{\alpha}J_{\Sigma} \, ,\\
\notag && J_{8,\alpha}=J_{\Sigma}^TC\gamma_5\gamma_{\alpha}J_{\Sigma} \, ,\\
&& J_{\Sigma}=\varepsilon^{ijk}\left(u^{iT}C\gamma_{\mu}d^j\right)\gamma^{\mu}\gamma_5c^k \, ,
\end{eqnarray}
 the superscripts $i, j, k$ are color indices and $C$ represents the charge conjugation matrix. The parities of these currents are shown as,
\begin{eqnarray}
\notag &&\widehat{P}J_{1,\alpha}(x)\widehat{P}^{-1}=J_1^\alpha(\widetilde{x}) \, ,\\
\notag &&\widehat{P}J_{2,\alpha}(x)\widehat{P}^{-1}=-J_2^\alpha(\widetilde{x}) \, ,\\
\notag &&\widehat{P}J_3(x)\widehat{P}^{-1}=J_3(\widetilde{x}) \, , \\
\notag &&\widehat{P}J_4(x)\widehat{P}^{-1}=-J_4(\widetilde{x}) \, , \\
\notag &&\widehat{P}J_5(x)\widehat{P}^{-1}=J_5(\widetilde{x}) \, , \\
\notag &&\widehat{P}J_6(x)\widehat{P}^{-1}=-J_6(\widetilde{x}) \, , \\
\notag &&\widehat{P}J_{7,\alpha}(x)\widehat{P}^{-1}=-J_7^\alpha(\widetilde{x}) \, ,\\
&&\widehat{P}J_{8,\alpha}(x)\widehat{P}^{-1}=J_8^\alpha(\widetilde{x}) \, ,
\end{eqnarray}

\noindent where $\widehat{P}$ is the parity operator, $x=(x^0,x^1,x^2,x^3)$ and $\widetilde{x}=(x^0,-x^1,-x^2,-x^3)$. A complete set of intermediate hadronic states are inserted into each $J_i(x)J^\dag_i(0)$ in the above correlation functions. The correlation functions at the hadron sides are simplified as,

\begin{eqnarray}
\notag &&\Pi_{1,2,7,8;\alpha\beta}(p)=A_{1,2,7,8}(p)\left(-g_{\alpha\beta}+\frac{p_\alpha p_\beta}{p^2}\right)+... \, ,\\
&& \Pi_{3,4,5,6}(p)=\frac{\lambda_{3,4,5,6}^2}{M_{3,4,5,6}^2-p^2}+... \, ,
\end{eqnarray}

\noindent where
\begin{eqnarray}
\notag && A_{1,2,7,8}(p)=\frac{\lambda_{1,2,7,8}^2}{M_{1,2,7,8}^2-p^2}+... \, ,\\
\notag &&\langle0 | J_{1,2,7,8;\alpha}(0)|Z_{1,2,7,8} \rangle =\lambda_{1,2,7,8}\epsilon_\alpha \, ,\\
&& \langle0 | J_{3,4,5,6}(0)|Z_{3,4,5,6} \rangle =\lambda_{3,4,5,6} \, ,
\end{eqnarray}

\noindent $|Z_{i}\rangle$ denote the ground states which have the same $J^P$ as $J_i$, the $\epsilon_\alpha$ are the polarization vectors and $\lambda_{i}$ are the pole residues which show the coupling of the currents $J_{i}$ to the states ($i=1,2,\cdot\cdot\cdot,8$).

For the QCD sides of the correlation functions, we contract the quark fields via the Wick theorem and express the correlation functions in terms of full quark propagators. After the detailed calculations, we find $\Pi_{7;\alpha\beta}(p)\equiv0$ which means the net contribution of all the Feynman diagrams of this state including the leading order, perturbation terms and all kinds of condensates is zero. Except for this term, the others are given by,
\begin{eqnarray}
\notag &&\Pi_{1;\alpha\beta}(p)=-\varepsilon^{ijk}\varepsilon^{lmn}\varepsilon^{i'j'k'}\varepsilon^{l'm'n'}i\int d^4x e^{ip\cdot x} \\
\notag && \Big\{Tr\left[\gamma_5\gamma^\mu\gamma_\alpha\gamma^\tau\gamma_5B^{nk'}(x)\gamma_5\gamma^{\mu'}\gamma_\beta\gamma^{\tau'}\gamma_5B^{n'k}(-x)\right] \\
\notag && Tr\left[\gamma_{\tau'}Q^{m'j}(-x)\gamma_\mu CQ^{l'iT}(-x)C\right]Tr\left[\gamma_\tau Q^{mj'}(x)\gamma_{\mu'}CQ^{li'T}(x)C\right]\Big\} \, ,
\end{eqnarray}
\begin{eqnarray}
\notag &&\Pi_{2;\alpha\beta}(p)=-\varepsilon^{ijk}\varepsilon^{lmn}\varepsilon^{i'j'k'}\varepsilon^{l'm'n'}i\int d^4x e^{ip\cdot x}   \\
\notag && \Big\{Tr\left[\gamma_5\gamma^\mu\gamma_5\gamma_\alpha\gamma^\tau\gamma_5B^{nk'}(x)\gamma_5\gamma^{\mu'}\gamma_5\gamma_\beta\gamma^{\tau'}\gamma_5B^{n'k}(-x)\right]\\
\notag && Tr\left[\gamma_{\tau'}Q^{m'j}(-x)\gamma_\mu CQ^{l'iT}(-x)C\right]Tr\left[\gamma_\tau Q^{mj'}(x)\gamma_{\mu'}CQ^{li'T}(x)C\right]\Big\} \, ,
\end{eqnarray}
\begin{eqnarray}
\notag &&\Pi_{3}(p)=-\varepsilon^{ijk}\varepsilon^{lmn}\varepsilon^{i'j'k'}\varepsilon^{l'm'n'}i\int d^4x e^{ip\cdot x}  \\
\notag && \Big\{Tr\left[\gamma_5\gamma^\mu\gamma^\tau\gamma_5B^{nk'}(x)\gamma_5\gamma^{\mu'}\gamma^{\tau'}\gamma_5B^{n'k}(-x)\right]\\
\notag && Tr\left[\gamma_{\tau'}Q^{m'j}(-x)\gamma_\mu CQ^{l'iT}(-x)C\right]Tr\left[\gamma_\tau Q^{mj'}(x)\gamma_{\mu'}CQ^{li'T}(x)C\right]\Big\} \, ,
\end{eqnarray}
\begin{eqnarray}
\notag &&\Pi_{4}(p)=-\varepsilon^{ijk}\varepsilon^{lmn}\varepsilon^{i'j'k'}\varepsilon^{l'm'n'}i\int d^4x e^{ip\cdot x}  \\
\notag && \Big\{Tr\left[\gamma_5\gamma^\mu\gamma_5\gamma^\tau\gamma_5B^{nk'}(x)\gamma_5\gamma^{\mu'}\gamma_5\gamma^{\tau'}\gamma_5B^{n'k}(-x)\right]\\
\notag && Tr\left[\gamma_{\tau'}Q^{m'j}(-x)\gamma_\mu CQ^{l'iT}(-x)C\right]Tr\left[\gamma_\tau Q^{mj'}(x)\gamma_{\mu'}CQ^{li'T}(x)C\right]\Big\} \, ,
\end{eqnarray}
\begin{eqnarray}
\notag &&\Pi_{5}(p)=-2\varepsilon^{ijk}\varepsilon^{lmn}\varepsilon^{i'j'k'}\varepsilon^{l'm'n'}i\int d^4x e^{ip\cdot x}  \\
\notag && \Big\{Tr\left[\gamma_5\gamma^{\tau}\gamma_5\gamma^\mu\gamma_5B^{kk'}(x)\gamma_5\gamma^{\mu'}\gamma_5\gamma^{\tau'}\gamma_5CB^{nn'T}(x)C\right]\\
\notag &&\ Tr\left[\gamma_\mu Q^{jj'}(x)\gamma_{\mu'}CQ^{ii'T}(x)C\right]Tr\left[\gamma_\tau Q^{mm'}(x)\gamma_{\tau'}CQ^{ll'T}(x)C\right] \\
\notag && +Tr\left[\gamma_5\gamma^{\tau}\gamma_5\gamma^\mu\gamma_5B^{kn'}(x)\gamma_5\gamma^{\mu'}\gamma_5\gamma^{\tau'}\gamma_5CB^{nk'T}(x)C\right]\\
\notag &&\ Tr\left[\gamma_\mu Q^{jj'}(x)\gamma_{\tau'}CQ^{ii'T}(x)C\right]Tr\left[\gamma_\tau Q^{mm'}(x)\gamma_{\mu'}CQ^{ll'T}(x)C\right] \\
\notag && -2Tr\left[\gamma_5\gamma^{\tau}\gamma_5\gamma^\mu\gamma_5B^{kk'}(x)\gamma_5\gamma^{\mu'}\gamma_5\gamma^{\tau'}\gamma_5CB^{nn'T}(x)C\right]\\
\notag &&\ Tr\left[\gamma_\mu Q^{jj'}(x)\gamma_{\mu'}CQ^{li'T}(x)C\gamma_\tau Q^{mm'}(x)\gamma_{\tau'}CQ^{il'T}(x)C\right]\Big\} \, ,
\end{eqnarray}
\begin{eqnarray}
\notag &&\Pi_{6}(p)=2\varepsilon^{ijk}\varepsilon^{lmn}\varepsilon^{i'j'k'}\varepsilon^{l'm'n'}i\int d^4x e^{ip\cdot x}  \\
\notag && \Big\{Tr\left[\gamma_5\gamma^{\tau}\gamma^\mu\gamma_5B^{kk'}(x)\gamma_5\gamma^{\mu'}\gamma^{\tau'}\gamma_5CB^{nn'T}(x)C\right]\\
\notag &&\ Tr\left[\gamma_\mu Q^{jj'}(x)\gamma_{\mu'}CQ^{ii'T}(x)C\right]Tr\left[\gamma_\tau Q^{mm'}(x)\gamma_{\tau'}CQ^{ll'T}(x)C\right] \\
\notag && +Tr\left[\gamma_5\gamma^{\tau}\gamma^\mu\gamma_5B^{kn'}(x)\gamma_5\gamma^{\mu'}\gamma^{\tau'}\gamma_5CB^{nk'T}(x)C\right]\\
\notag &&\ Tr\left[\gamma_\mu Q^{jj'}(x)\gamma_{\tau'}CQ^{ii'T}(x)C\right]Tr\left[\gamma_\tau Q^{mm'}(x)\gamma_{\mu'}CQ^{ll'T}(x)C\right] \\
\notag && -2Tr\left[\gamma_5\gamma^{\tau}\gamma^\mu\gamma_5B^{kk'}(x)\gamma_5\gamma^{\mu'}\gamma^{\tau'}\gamma_5CB^{nn'T}(x)C\right]\\
\notag &&\ Tr\left[\gamma_\mu Q^{jj'}(x)\gamma_{\mu'}CQ^{li'T}(x)C\gamma_\tau Q^{mm'}(x)\gamma_{\tau'}CQ^{il'T}(x)C\right]\Big\} \, ,
\end{eqnarray}
\begin{eqnarray}
\notag &&\Pi_{8;\alpha\beta}(p)=2\varepsilon^{ijk}\varepsilon^{lmn}\varepsilon^{i'j'k'}\varepsilon^{l'm'n'}i\int d^4x e^{ip\cdot x}  \\
\notag && \Big\{Tr\left[\gamma_5\gamma^{\tau}\gamma_5\gamma_{\alpha}\gamma^\mu\gamma_5B^{kk'}(x)\gamma_5\gamma^{\mu'}\gamma_5\gamma_{\beta}\gamma^{\tau'}\gamma_5CB^{nn'T}(x)C\right]\\
\notag &&\ Tr\left[\gamma_\mu Q^{jj'}(x)\gamma_{\mu'}CQ^{ii'T}(x)C\right]Tr\left[\gamma_\tau Q^{mm'}(x)\gamma_{\tau'}CQ^{ll'T}(x)C\right]\\
\notag && +Tr\left[\gamma_5\gamma^{\tau}\gamma_5\gamma_{\alpha}\gamma^\mu\gamma_5B^{kn'}(x)\gamma_5\gamma^{\mu'}\gamma_5\gamma_{\beta}\gamma^{\tau'}\gamma_5CB^{nk'T}(x)C\right]\\
\notag &&\ Tr\left[\gamma_\mu Q^{jj'}(x)\gamma_{\tau'}CQ^{ii'T}(x)C\right]Tr\left[\gamma_\tau Q^{mm'}(x)\gamma_{\mu'}CQ^{ll'T}(x)C\right]\\
\notag && -2Tr\left[\gamma_5\gamma^{\tau}\gamma_5\gamma_{\alpha}\gamma^\mu\gamma_5B^{kk'}(x)\gamma_5\gamma^{\mu'}\gamma_5\gamma_{\beta}\gamma^{\tau'}\gamma_5CB^{nn'T}(x)C\right]\\
 &&\ Tr\left[\gamma_\mu Q^{jj'}(x)\gamma_{\mu'}CQ^{li'T}(x)C\gamma_\tau Q^{mm'}(x)\gamma_{\tau'}CQ^{il'T}(x)C\right]\Big\} \, ,
\end{eqnarray}

\noindent where $Q^{ab}(x)$ and $B^{ab}(x)$ are the full light and heavy quark propagators, respectively, under the operator product expansion at the quark level, they are expressed as,
\begin{eqnarray}
\notag\ Q^{ab}(x)=&& \frac{ix\!\!\!/\delta^{ab}}{2\pi^{2}x^{4}}-\frac{\delta^{ab}}{12}\langle\overline{q}q\rangle-\frac{\delta^{ab}x^2}{192}\langle\overline{q}g_s\sigma G q\rangle-\frac{i\delta^{ab}x^2x\!\!\!/g_s^2\langle\overline{q}q\rangle^2}{7776}-\left(t^n\right)^{ab}\left(x\!\!\!/\sigma^{\alpha\beta}+\sigma^{\alpha\beta}x\!\!\!/\right)\frac{i}{32\pi^2x^2}g_s G_{\alpha\beta}^n  \\
\notag\ &&-\frac{\delta^{ab}x^4\langle\overline{q}q\rangle\langle GG \rangle}{27648}-\frac{1}{8}\langle\overline{q}^b\sigma^{\alpha\beta}q^a\rangle\sigma_{\alpha\beta}-\frac{1}{4}\langle\overline{q}^b\gamma_\mu q^a\rangle\gamma^\mu+\cdot\cdot\cdot\, ,
\end{eqnarray}

\begin{eqnarray}
\notag\
B_{ab}(x)&&=\frac{i}{(2\pi)^{4}}\int d^{4}ke^{-ik\cdot x}\bigg\{\frac{\delta_{ab}}{k\!\!\!/-m_{c}}-\frac{g_{s}G_{\alpha\beta }^{h}t_{ab}^{h}}{4}\frac{\sigma^{\alpha\beta}(k\!\!\!/+m_{c})+(k\!\!\!/+m_{c})\sigma ^{\alpha
\beta }}{(k^{2}-m_{c}^{2})^{2}}\\
\notag\
&&+\frac{g_{s}D_{\alpha}G_{\beta\lambda}^{h}t_{ab}^{h}\left(f^{\lambda\beta\alpha}+f^{\lambda\alpha\beta}\right)}{3(k^{2}-m_{c}^{2})^{4}}
+\cdot\cdot\cdot\bigg \}\, ,
\end{eqnarray}

\begin{eqnarray}
 &&f^{\lambda \alpha \beta }=(k\!\!\!/+m_{c})\gamma ^{\lambda
}(k\!\!\!/+m_{c})\gamma ^{\alpha }(k\!\!\!/+m_{c})\gamma ^{\beta
}(k\!\!\!/+m_{c})\, ,
\end{eqnarray}

\noindent where $t^n=\frac{\lambda^n}{2}$, $\lambda^n$ are the Gell-Mann matrices ($n=1,2,\cdot\cdot\cdot,8$) and $D_\alpha=\partial_\alpha-ig_sG_\alpha^ht^h$ \cite{Reinders,Wang3,Pascual}.

Consider the Feynman graphes of the leading orders of these non-zero correlation functions, each one contains two heavy quark lines and four light quark lines. If each heavy quark line emits a gluon and each light quark line contributes quark-antiquark pair, they form the quark-gluon operator $g_sG_{\alpha\beta}g_sG_{\eta\tau}\overline{q}q\overline{q}q\overline{q}q\overline{q}q$ with dimension 16. This operator lead to vacuum condensates $\langle\frac{\alpha_s}{\pi}GG\rangle\langle\overline{q}q\rangle^4$ and $\langle\overline{q}g_s\sigma Gq\rangle^2\langle\overline{q}q\rangle^2$. So, the highest dimension of the considered condensates is up to 16 in the present work. As for the truncation of the order $\mathcal{O}(\alpha_s^k )$, the contributions of the related vacuum condensates are tiny while $k\geq\frac{3}{2}$ , it is accurate enough for us to calculate terms for $k\leq1$ \cite{wangxiuwu}. Detailed calculations show that the vacuum condensates $\langle\overline{q}q\rangle$ and $\langle\overline{q}g_s\sigma Gq\rangle$ have no contribution for all the considered states in this paper. Thus, the chosen vacuum condensates of the operator product expansions are $\langle\frac{\alpha_s}{\pi}GG\rangle$, $\langle\overline{q}q\rangle^2$, $\langle\overline{q}g_s\sigma Gq\rangle\langle\overline{q}q\rangle$, $\langle\overline{q}g_s\sigma Gq\rangle^2$, $\langle\frac{\alpha_s}{\pi}GG\rangle\langle\overline{q}q\rangle^2$, $\langle\overline{q}q\rangle^4$, $\langle\overline{q}g_s\sigma Gq\rangle\langle\overline{q}q\rangle^3$, $\langle\overline{q}g_s\sigma Gq\rangle^2\langle\overline{q}q\rangle^2$ and $\langle\frac{\alpha_s}{\pi}GG\rangle\langle\overline{q}q\rangle^4$.

After the calculations of the integrals in the coordinate space for the light quarks and momentum space for heavy quarks of the correlation functions, we perform the Borel transforms for both the hadronic sides and the QCD sides, we get the QCD sum rules,
\begin{eqnarray}
\lambda_{i}^2\exp\left(-\frac{M_{i}^2}{T^2}\right)=\int_{\Delta^2}^{s_0}ds \rho_{i}(s)\exp\left(-\frac{s}{T^2}\right)\, ,
\end{eqnarray}

\noindent where $\rho_{i}(s)$ are the QCD spectral densities of $J_{i}$, $\Delta^2=4m_c^2$. As for the continuum threshold parameters $s_0$, we take the experiential values \cite{wangxiuwu,WZG,XZWZG2},
\begin{eqnarray}
\sqrt{s_0}=M_{Z}+(0.5\sim0.7){\rm GeV}\, ,
\end{eqnarray}
where $M_{Z}$ are the masses of the ground states.

We differentiate the Eq.(8) with respect to $\tau=\frac{1}{T^2}$ and eliminate the pole residues, the extracted masses are shown as,
\begin{eqnarray}
M_{i}^2=\frac{-\frac{d}{d\tau}\int_{\Delta^2}^{s_0}ds \rho_{i}(s)\exp(-s\tau)}{\int_{\Delta^2}^{s_0}ds \rho_{i}(s)\exp(-s\tau)}\, .
\end{eqnarray}

\begin{large}
\noindent\textbf{3 Numerical results and discussions}
\end{large}

We apply the standard values of the vacuum condensates $\langle\overline{q}q\rangle=-(0.24\pm0.01\;{\rm GeV})^3$, $\langle\overline{q}g_s\sigma Gq\rangle=m_0^2\langle\overline{q}q\rangle\;$GeV$^2$, $m_0^2=(0.8\pm0.1)\;{\rm GeV}^2$, $\langle\frac{\alpha_s}{\pi}GG\rangle=(0.33\;{\rm GeV})^4$ at the energy scale $\mu=1\;{\rm GeV}$ \cite{Reinders,Patrignani}, and choose the $
\overline{MS}$ mass $m_c(m_c)=1.275\pm0.025\;{\rm GeV}$ \cite{PDG}. Consider the energy-scale dependence for the above condensates,
\begin{eqnarray}
\notag \langle\overline{q}q\rangle(\mu)&&=\langle\overline{q}q\rangle(1\;{\rm GeV})\left[\frac{\alpha_s(1\;{\rm GeV})}{\alpha_s(\mu)}\right]^{\frac{12}{33-2n_f}}\, ,\\
\notag \langle\overline{q}g_s\sigma Gq\rangle(\mu)&& =\langle\overline{q}g_s\sigma Gq\rangle(1\;{\rm GeV})\left[\frac{\alpha_s(1\;{\rm GeV})}{\alpha_s(\mu)}\right]^{\frac{2}{33-2n_f}}\, ,\\
\notag  m_c(\mu)&&=m_c(m_c)\left[\frac{\alpha_s(\mu)}{\alpha_s(m_c)}\right]^{\frac{12}{33-2n_f}}\, ,\\
\notag \alpha_s(\mu)&&=\frac{1}{b_0t}\left[1-\frac{b_1}{b_0^2}\frac{\rm{log}\emph{t}}{t}+\frac{b_1^2(\rm{log}^2\emph{t}-\rm{log}\emph {t}-1)+\emph{b}_0\emph{b}_2}{b_0^4t^2}\right]\, ,
\end{eqnarray}
where $t=\rm{log}\frac{\mu^2}{\Lambda_{\emph{QCD}}^2}$, $\emph b_0=\frac{33-2\emph{n}_\emph{f}}{12\pi}$, $b_1=\frac{153-19n_f}{24\pi^2}$, $b_2=\frac{2857-\frac{5033}{9}n_f+\frac{325}{27}n_f^2}{128\pi^3}$
and $\Lambda_{QCD}=213$ MeV, $296$ MeV, $339$ MeV for the flavors $n_f=5,4,3$, respectively \cite{Narison,PDG,Shifman}, In this paper, we choose the flavor number $n_f=4$ for all the states. Compared with the heavy quarks, the masses of the light quarks are too small to make effective difference. So, we ignore the masses of the light quarks. The masses of the baryon and anti-baryon $M_{\Sigma_c}=M_{\overline\Sigma_c}=2.455$ {\rm GeV} are from the Particle Data Group \cite{PDG}. We abandon the expressions of the lengthy spectral densities of these currents which are dozens of pages, one can contact us via the email.
Note that, Eq.(10) depends on the energy scale $\mu$ \cite{Chen,Wang4}, we apply the energy scale formula to determine the best energy scales of the QCD spectral densities \cite{Wang3},
\begin{eqnarray}
\mu=\sqrt{M_{X/Y/Z}^2-4\mathbb{M}_c^2}\, ,
\end{eqnarray}
where $\mathbb{M}_c$ is the effective charm quark mass, we choose the value $\mathbb{M}_c=1.84\pm0.01$ \rm GeV \cite{wangzggg}. Interestingly, in Ref. \cite{wangxiuwu}, for the vector current $J_\mu=\bar{J}_{\Lambda}\gamma_{\mu}J_{\Lambda}$, if we set $\mathbb{M}_c=1.84$ GeV, the central value of the mass of $\Lambda_c\bar{\Lambda}_c$ with the $J^P=1^-$ is $4.57$ GeV with $\sqrt{s_0}=2m_\Lambda+0.66$ GeV and $\rm{PC}=39\%-53\%$ which indicates a possible bound $\Lambda_c\bar{\Lambda}_c$ baryonium state, where $m_\Lambda=2.29$ GeV is the mass of $\Lambda_c$ \cite{PDG}. As part of the systematic study for the dibaryons formed by the charmed baryons, the detailed results of Ref. \cite{wangxiuwu} are also shown in the Table \uppercase\expandafter{\romannumeral1}.
From Eq.(10), we find that $M_{Z}$ relies on the threshold parameter $s_0$ and Borel parameter $T^2$. Of course, at first, we do not know the suitable energy scale, so we try and adjust the value of $\mu$ via trial and error. From the energy scale formula Eq.(11), one can determine that $M_Z$ increases with the parameter $\mu$, however, numerical results show that mass derived from Eq.(10) decreases with the increment of the parameter $\mu$, so we can always find the best energy scale and determine the mass which satisfies the energy scale formula. As for the values of $s_0$ and $T^2$, we search for the best Borel window which satisfies the following two principles, firstly, the pole dominance criterion, here we define the pole contributions (PC) of the QCD sum rules as,
\begin{eqnarray}
{\rm PC}=\frac{\int_{4m_c^2}^{s_0}ds\rho_{QCD}(s)\exp\left(-\frac{s}{T^2}\right)}{\int_{4m_c^2}^{\infty}ds\rho_{QCD}(s)\exp\left(-\frac{s}{T^2}\right)}\, ,
\end{eqnarray}
\noindent secondly, the convergence of the contribution of operator product expansion, for which, we define the contributions of the vacuum condensates of dimension $n$ as,
\begin{eqnarray}
D(n)=\frac{\int_{4m_c^2}^{s_0}ds\rho_{QCD;n}(s)\exp\left(-\frac{s}{T^2}\right)}{\int_{4m_c^2}^{s_0}ds\rho_{QCD}(s)\exp\left(-\frac{s}{T^2}\right)}\, ,
\end{eqnarray}

\begin{figure}[htpb]
\centering
\subfigure{
\begin{minipage}[h]{4cm}
\centering
\includegraphics[height=4cm,width=4.5cm]{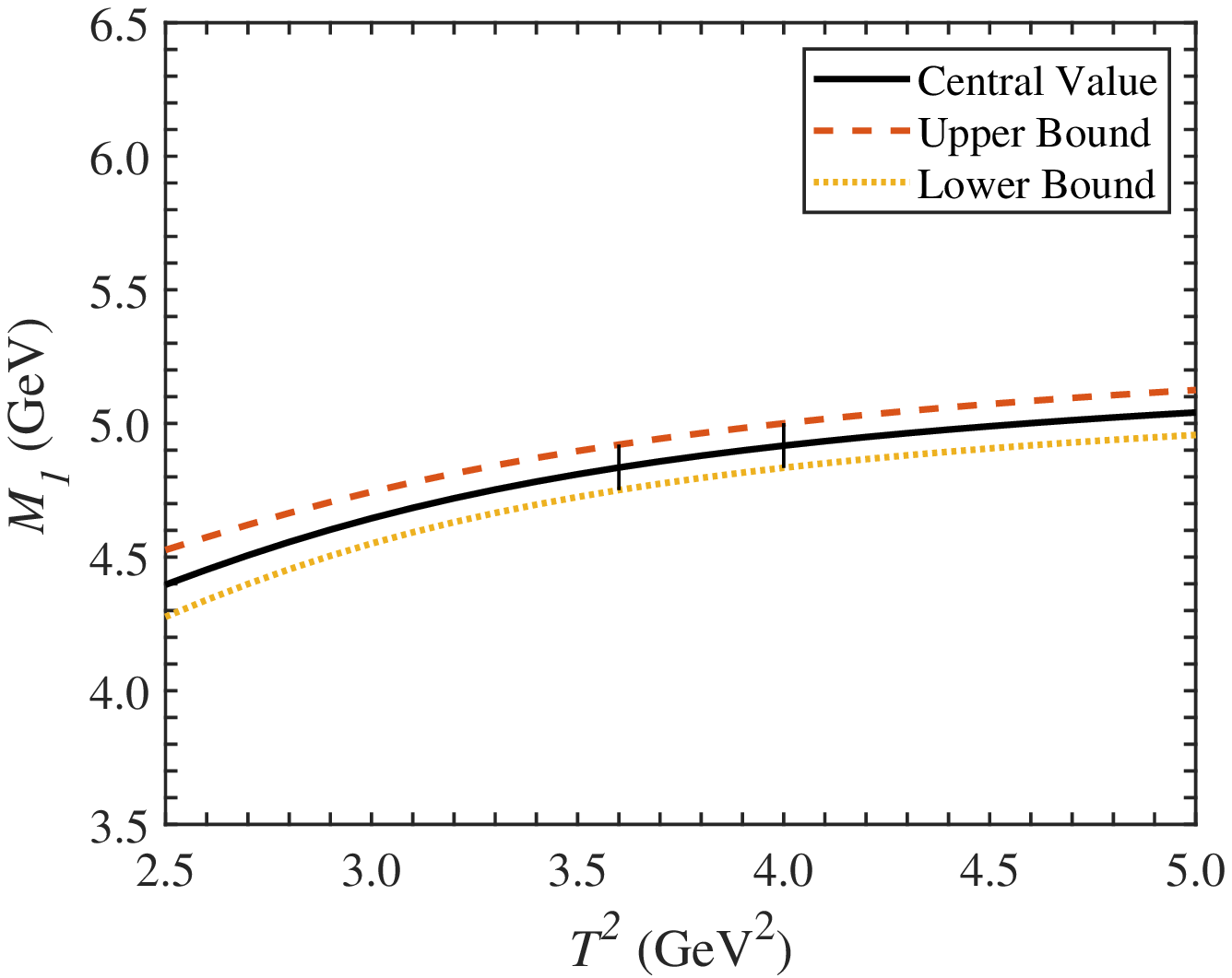}
\end{minipage}
}
\subfigure{
\begin{minipage}[h]{4cm}
\centering
\includegraphics[height=4cm,width=4.5cm]{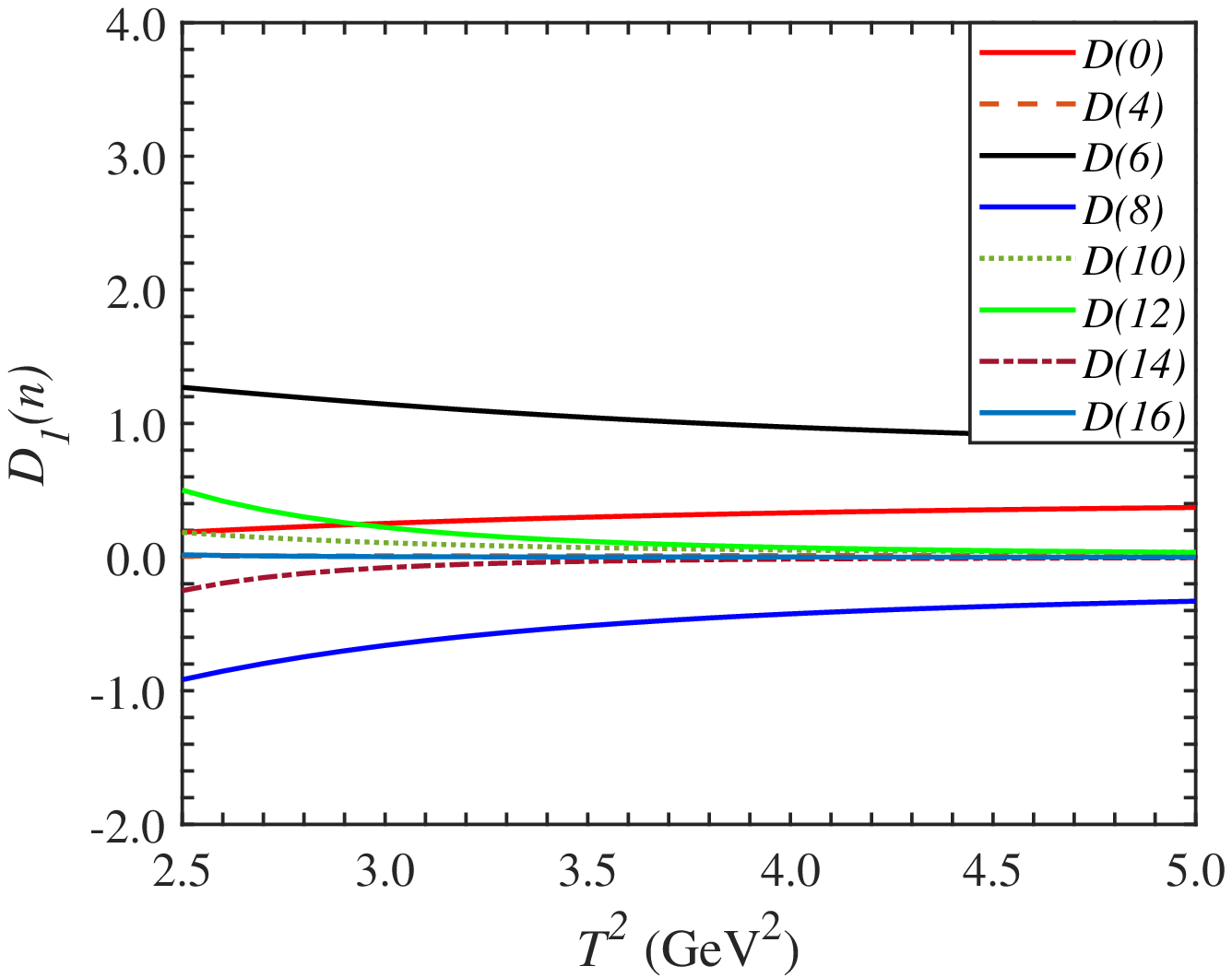}
\end{minipage}
}
\subfigure{
\begin{minipage}[h]{4cm}
\centering
\includegraphics[height=4cm,width=4.5cm]{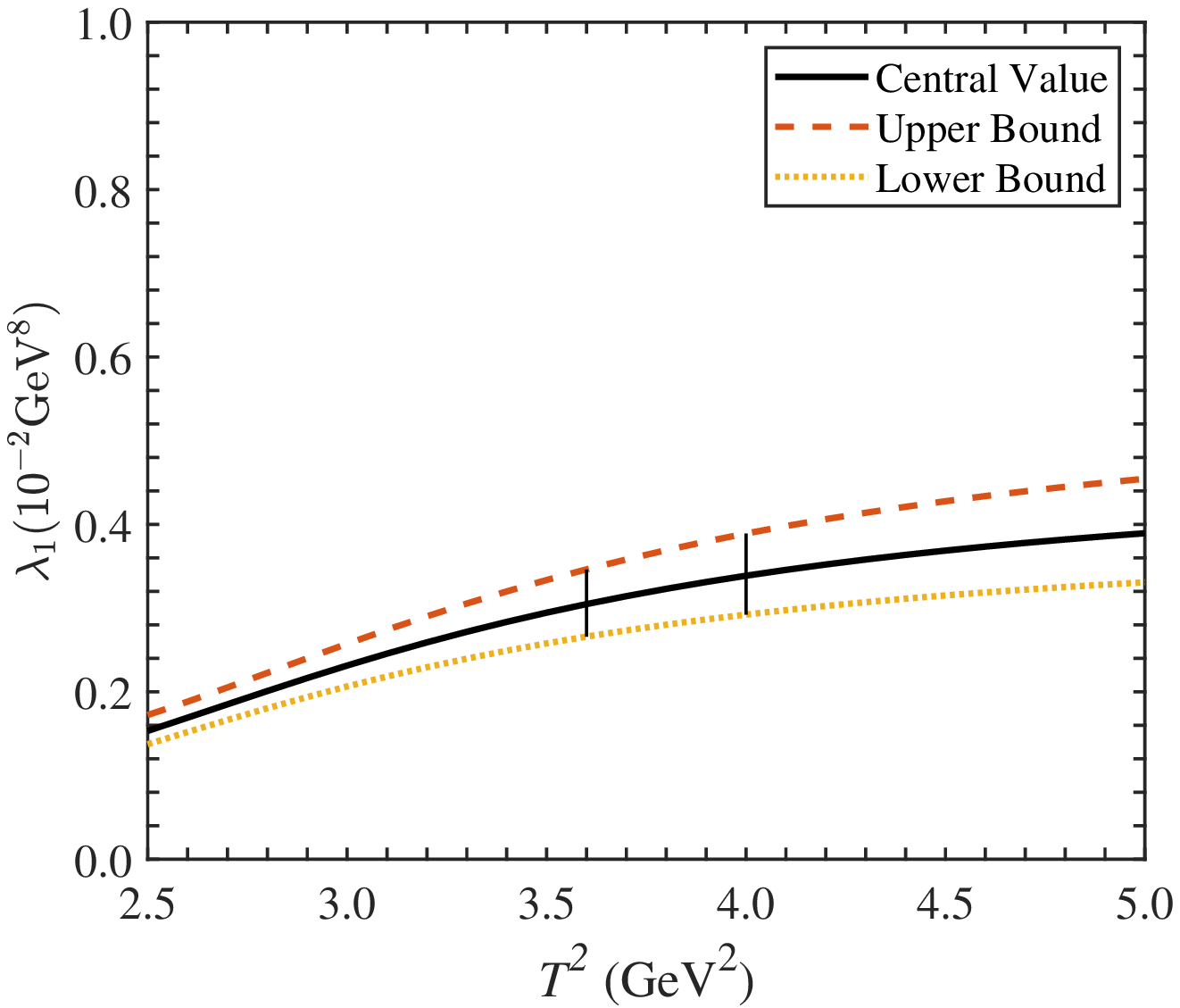}
\end{minipage}
}
\caption{The numerical results of $J_1$: $M_1-T^2$ for the left graph, dashed curve is drawn from the upper values of the input parameters and dotted curve is from the lower values; $D_1(n)-T^2$ for the middle graph; $\lambda_1-T^2$ for the right graph, dashed curve is drawn from the upper values of the input parameters and dotted curve is from the lower values. \label{your label}}
\end{figure}

\begin{figure}[htpb]
\centering
\subfigure{
\begin{minipage}[h]{4cm}
\centering
\includegraphics[height=4cm,width=4.5cm]{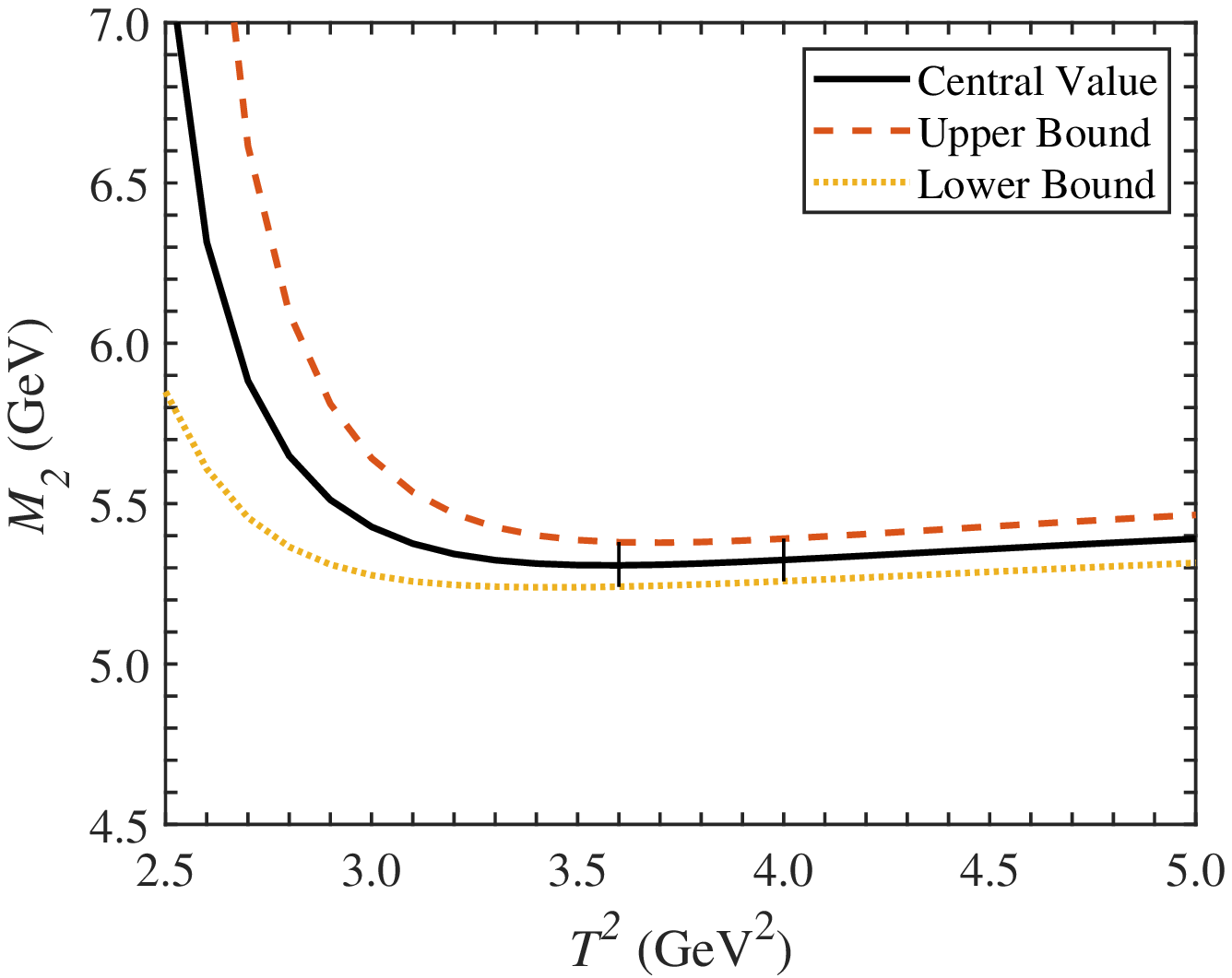}
\end{minipage}
}
\subfigure{
\begin{minipage}[h]{4cm}
\centering
\includegraphics[height=4cm,width=4.5cm]{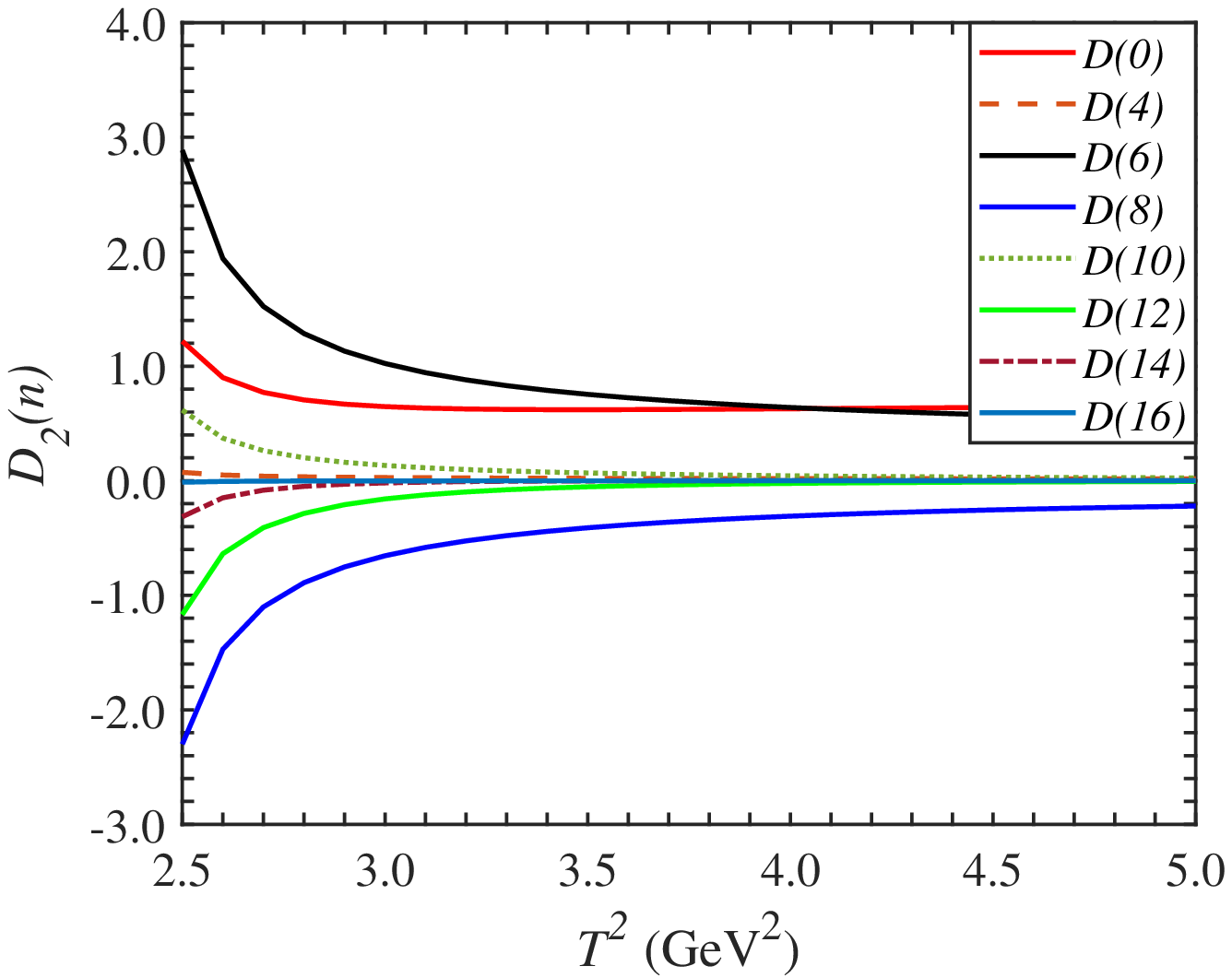}
\end{minipage}
}
\subfigure{
\begin{minipage}[h]{4cm}
\centering
\includegraphics[height=4cm,width=4.5cm]{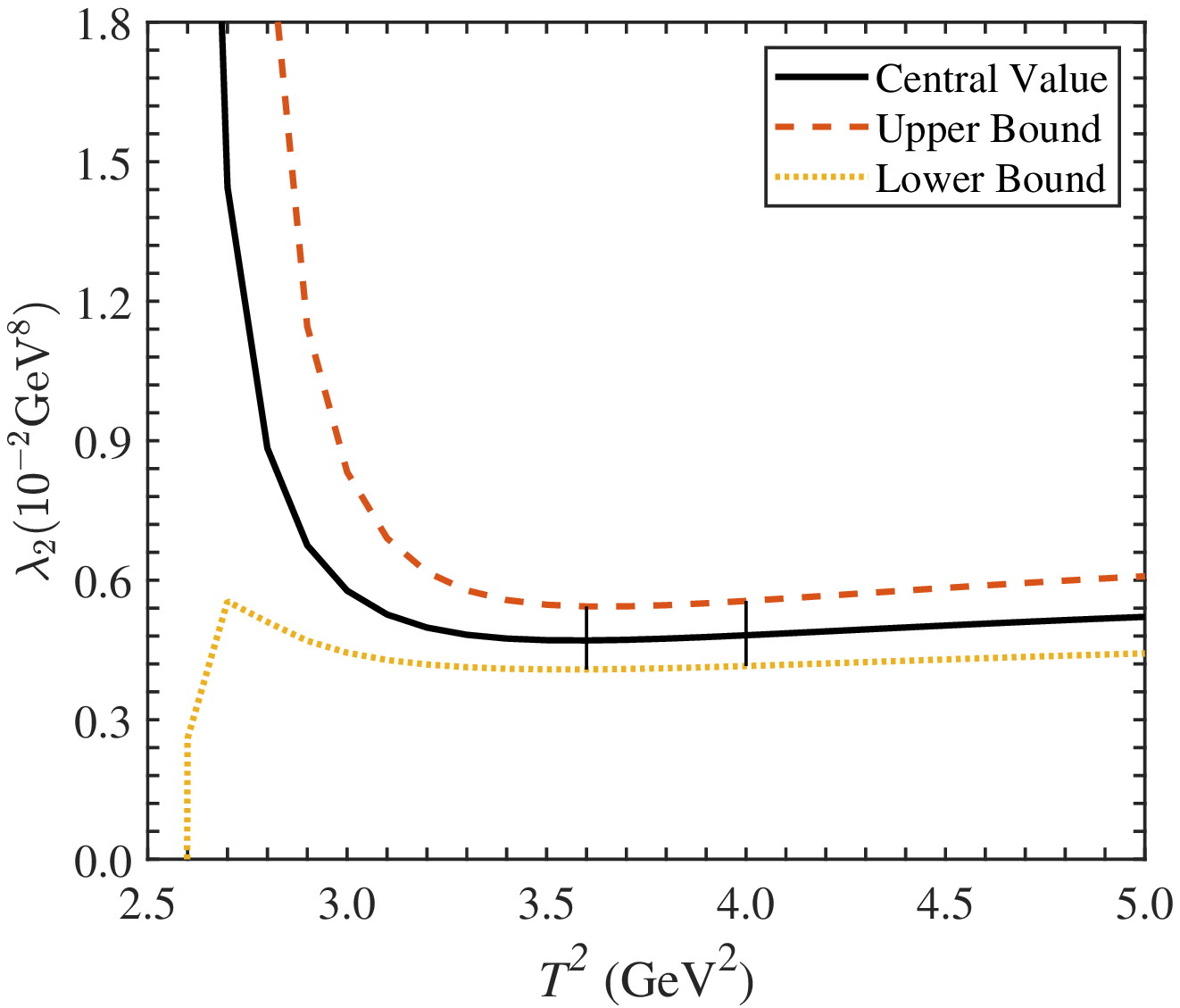}
\end{minipage}
}
\caption{The numerical results of $J_2$: $M_2-T^2$ for the left graph, dashed curve is drawn from the upper values of the input parameters and dotted curve is from the lower values; $D_2(n)-T^2$ for the middle graph; $\lambda_2-T^2$ for the right graph, dashed curve is drawn from the upper values of the input parameters and dotted curve is from the lower values. \label{your label}}
\end{figure}

\begin{figure}[htpb]
\centering
\subfigure{
\begin{minipage}[h]{4cm}
\centering
\includegraphics[height=4cm,width=4.5cm]{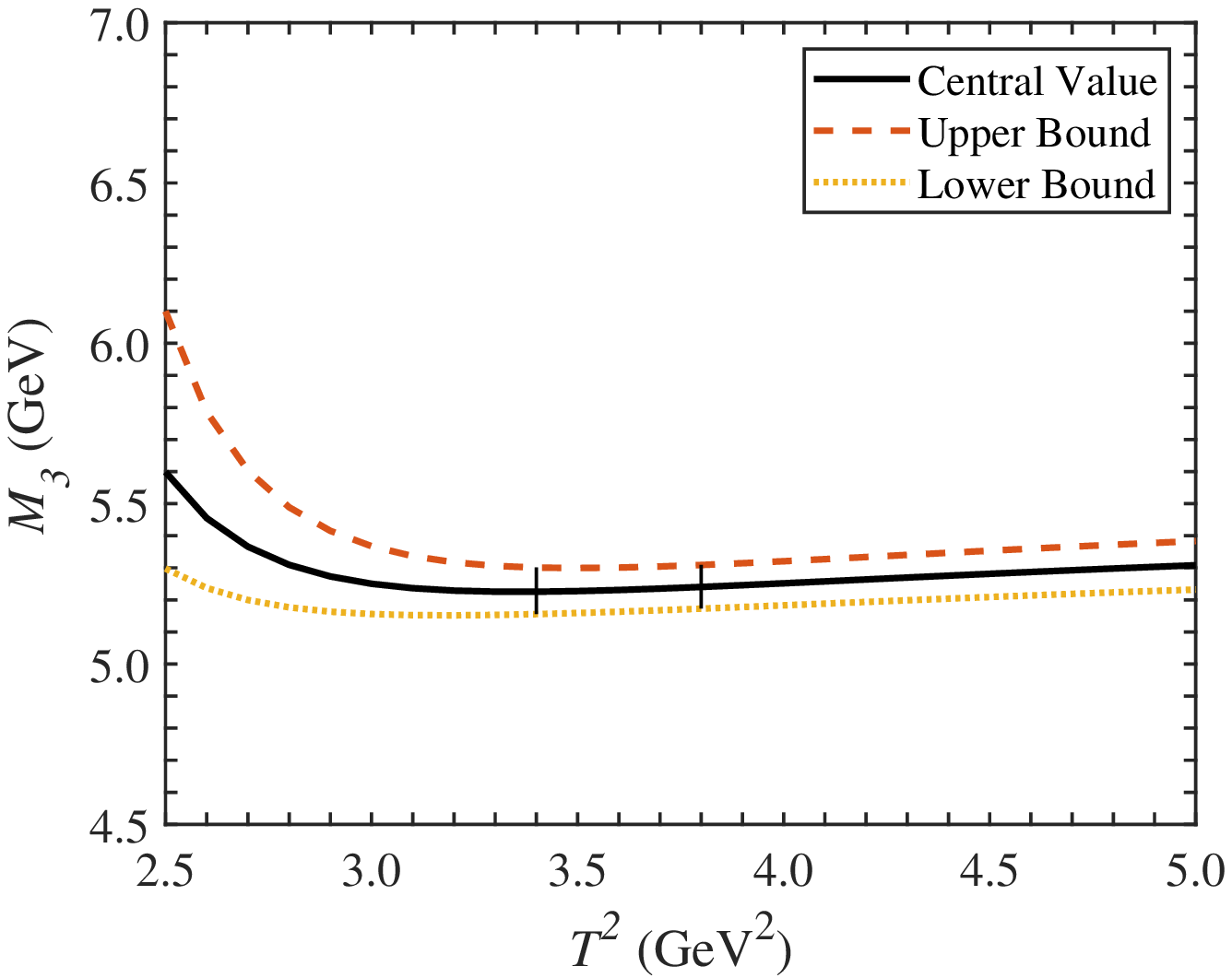}
\end{minipage}
}
\subfigure{
\begin{minipage}[h]{4cm}
\centering
\includegraphics[height=4cm,width=4.5cm]{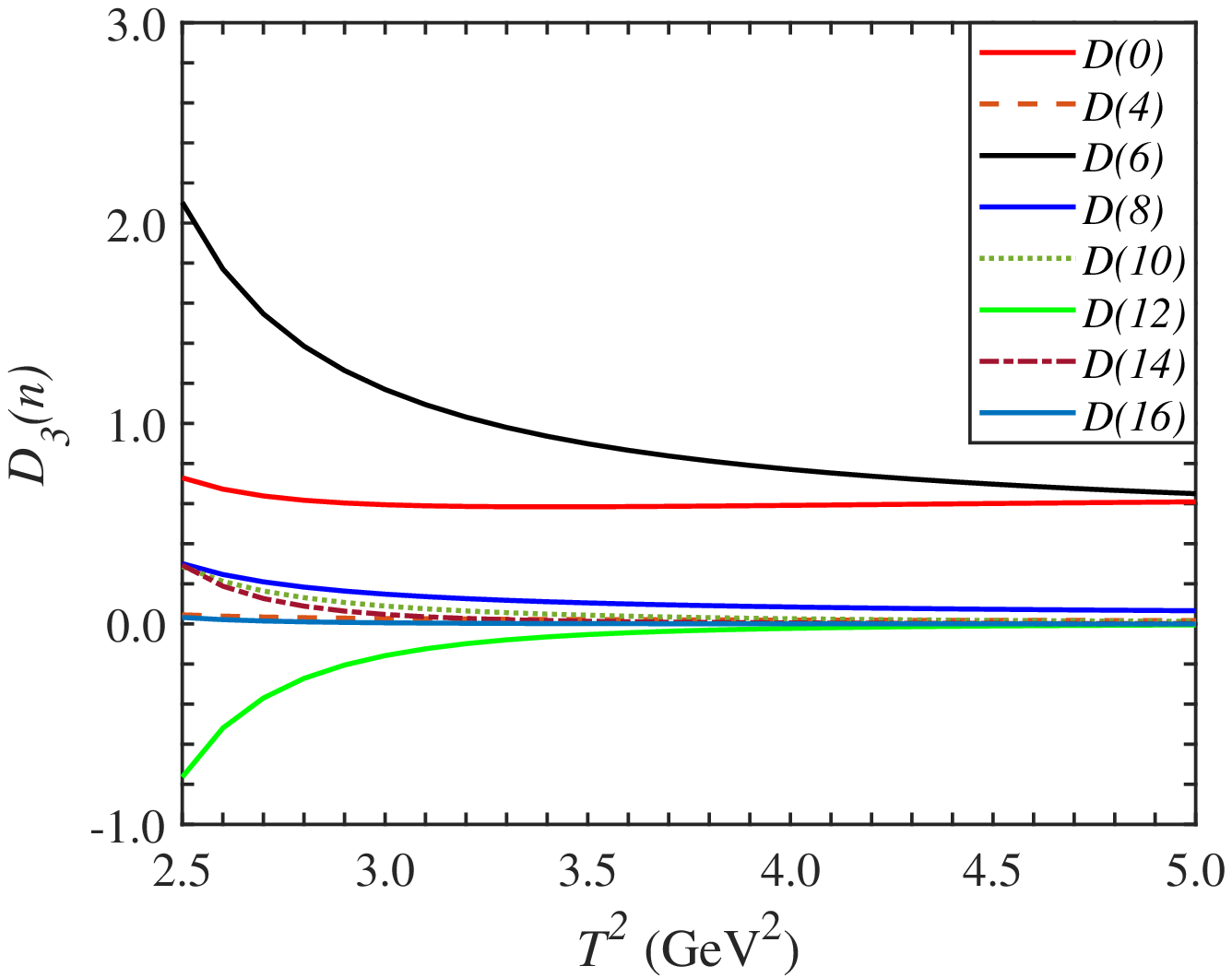}
\end{minipage}
}
\subfigure{
\begin{minipage}[h]{4cm}
\centering
\includegraphics[height=4cm,width=4.5cm]{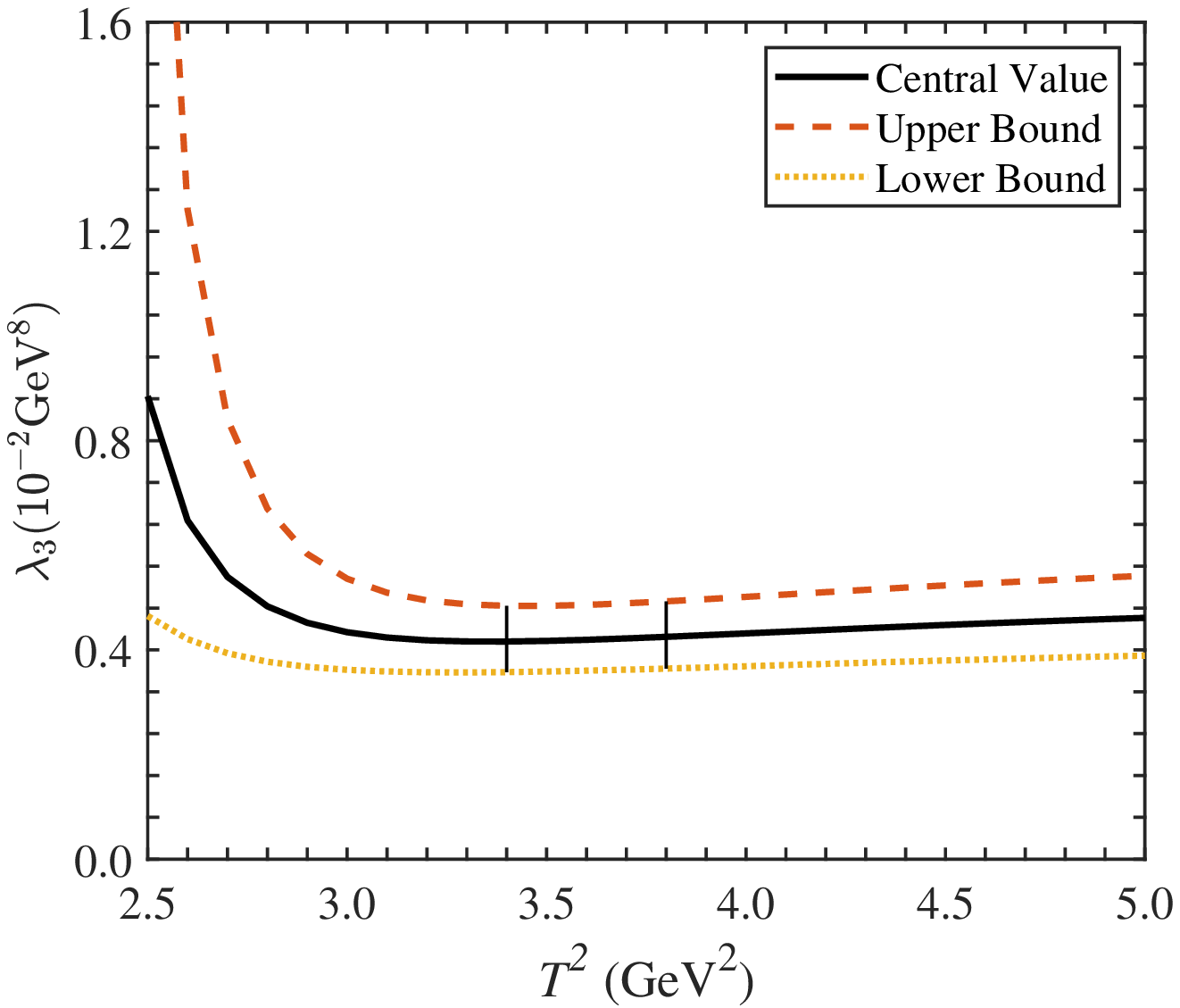}
\end{minipage}
}
\caption{The numerical results of $J_3$: $M_3-T^2$ for the left graph, dashed curve is drawn from the upper values of the input parameters and dotted curve is from the lower values; $D_3(n)-T^2$ for the middle graph; $\lambda_3-T^2$ for the right graph, dashed curve is drawn from the upper values of the input parameters and dotted curve is from the lower values. \label{your label}}
\end{figure}

\begin{figure}[htpb]
\centering
\subfigure{
\begin{minipage}[h]{4cm}
\centering
\includegraphics[height=4cm,width=4.5cm]{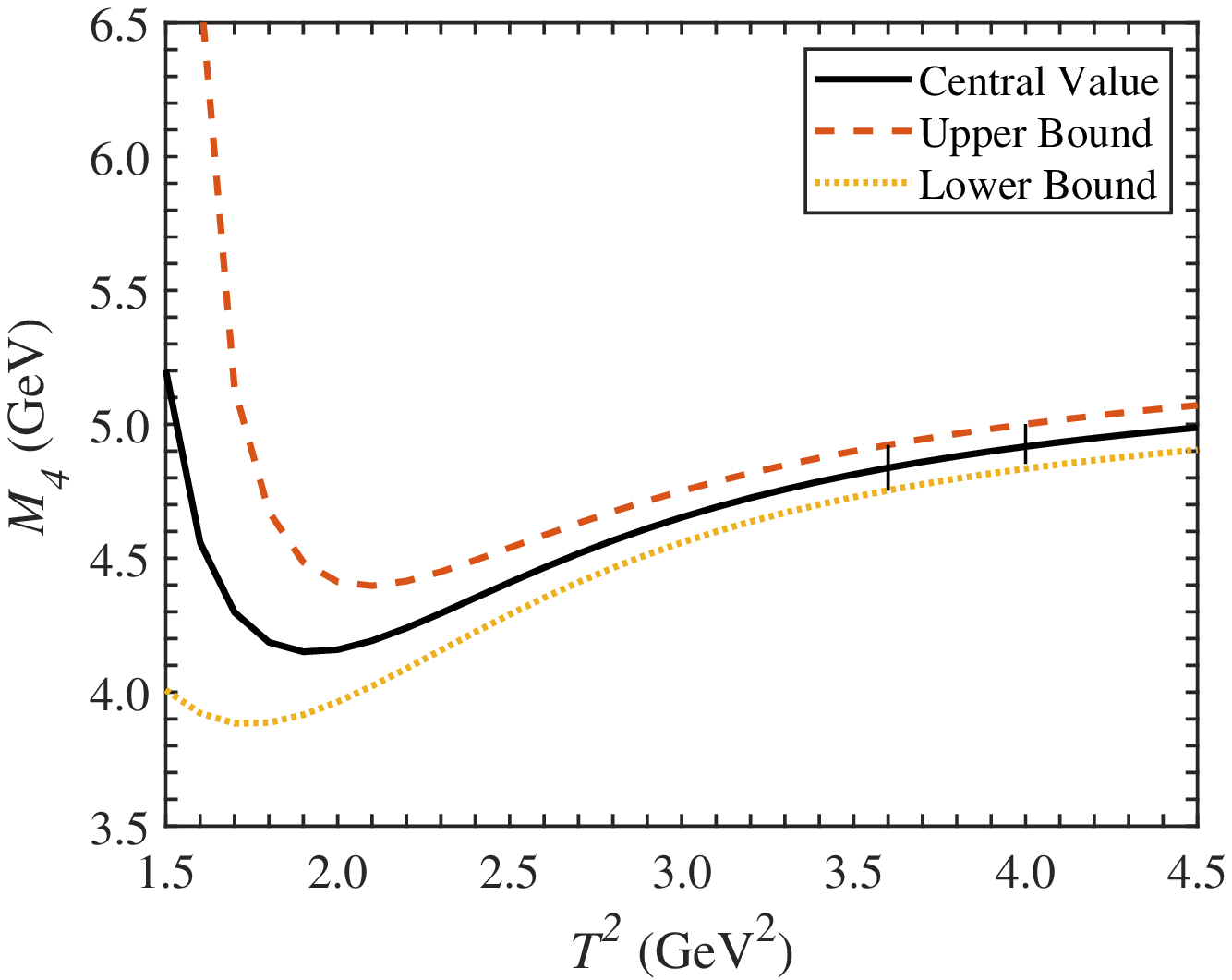}
\end{minipage}
}
\subfigure{
\begin{minipage}[h]{4cm}
\centering
\includegraphics[height=4cm,width=4.5cm]{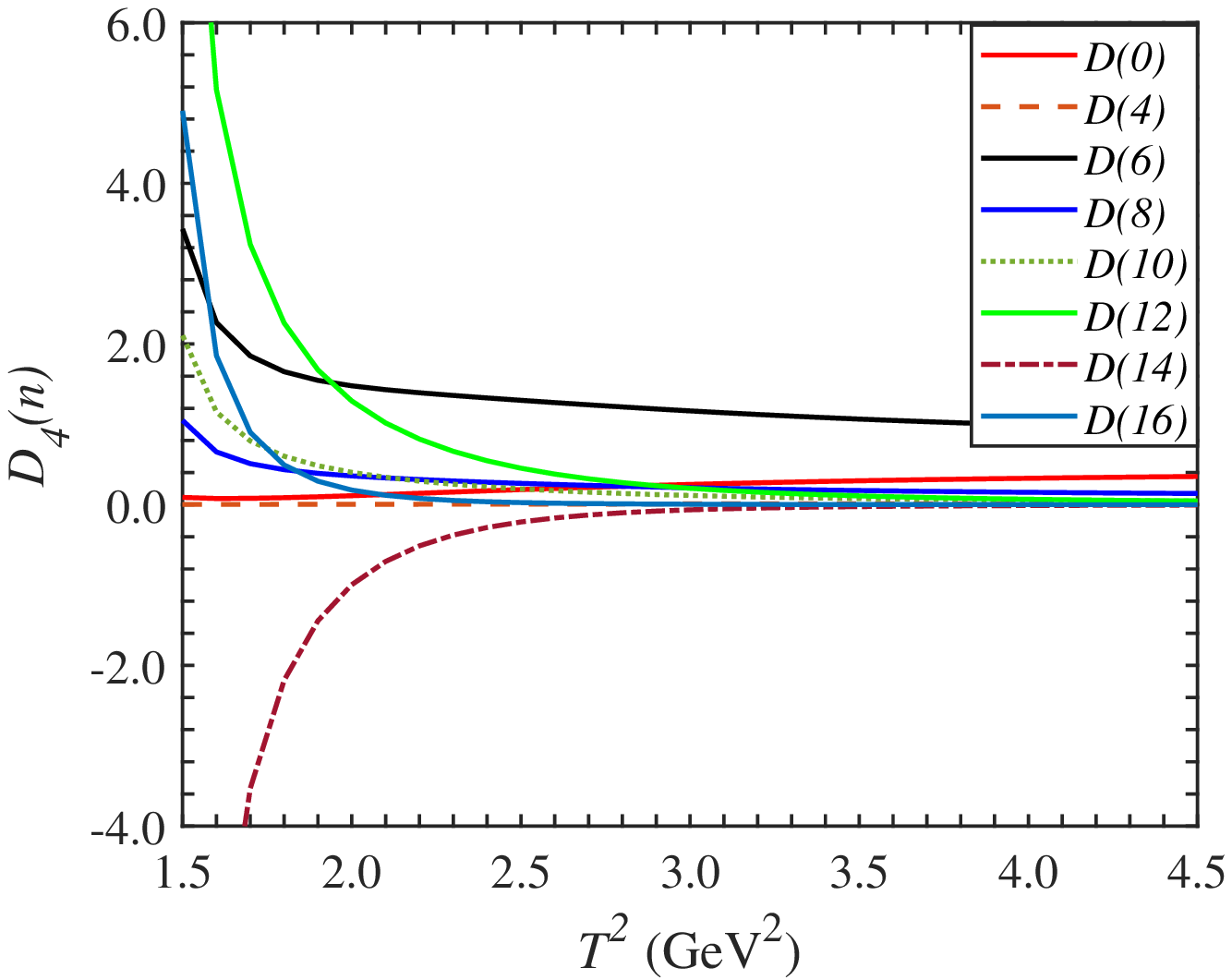}
\end{minipage}
}
\subfigure{
\begin{minipage}[h]{4cm}
\centering
\includegraphics[height=4cm,width=4.5cm]{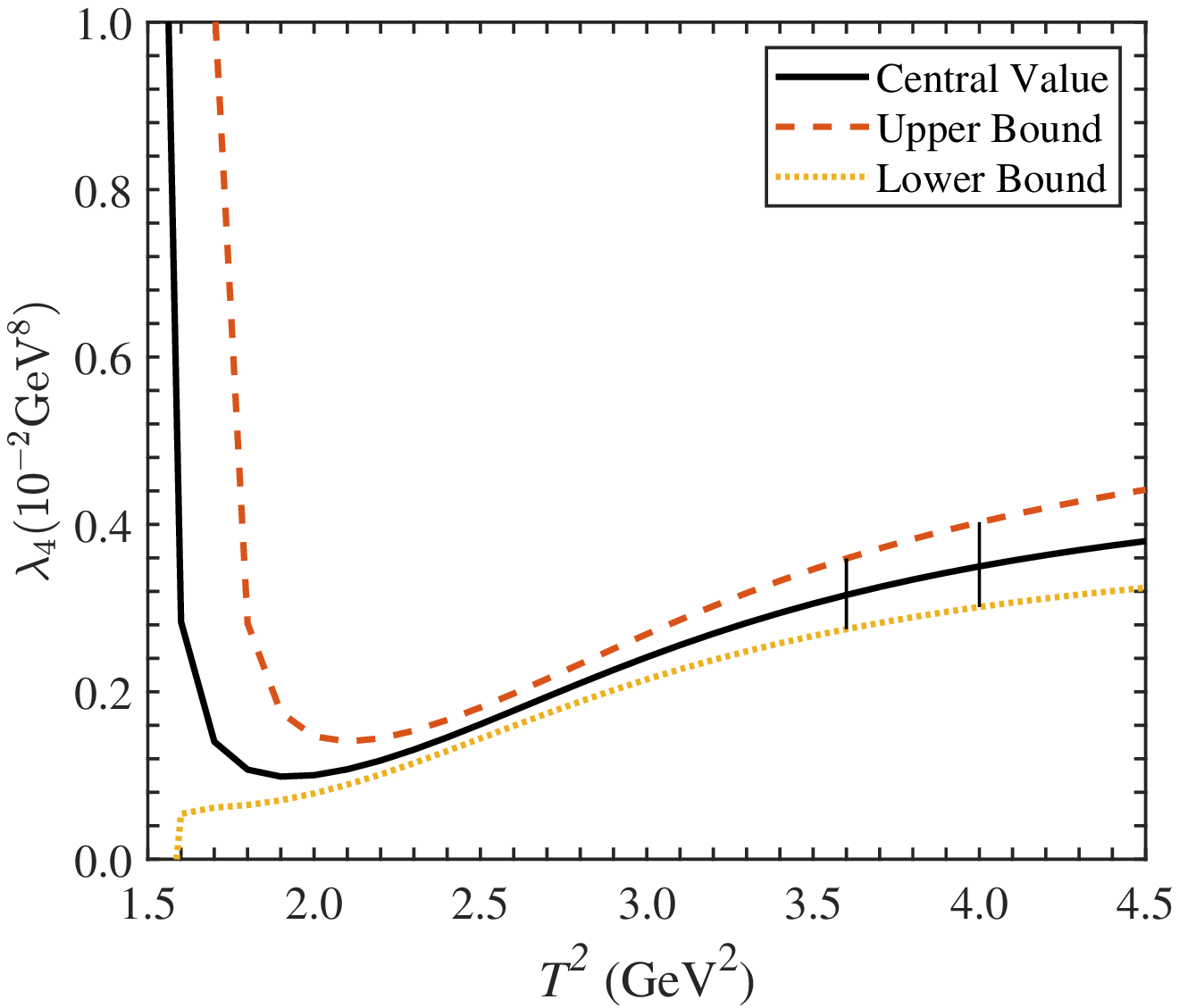}
\end{minipage}
}
\caption{The numerical results of $J_4$: $M_4-T^2$ for the left graph, dashed curve is drawn from the upper values of the input parameters and dotted curve is from the lower values; $D_4(n)-T^2$ for the middle graph; $\lambda_4-T^2$ for the right graph, dashed curve is drawn from the upper values of the input parameters and dotted curve is from the lower values. \label{your label}}
\end{figure}

\begin{figure}[htpb]
\centering
\subfigure{
\begin{minipage}[h]{4cm}
\centering
\includegraphics[height=4cm,width=4.5cm]{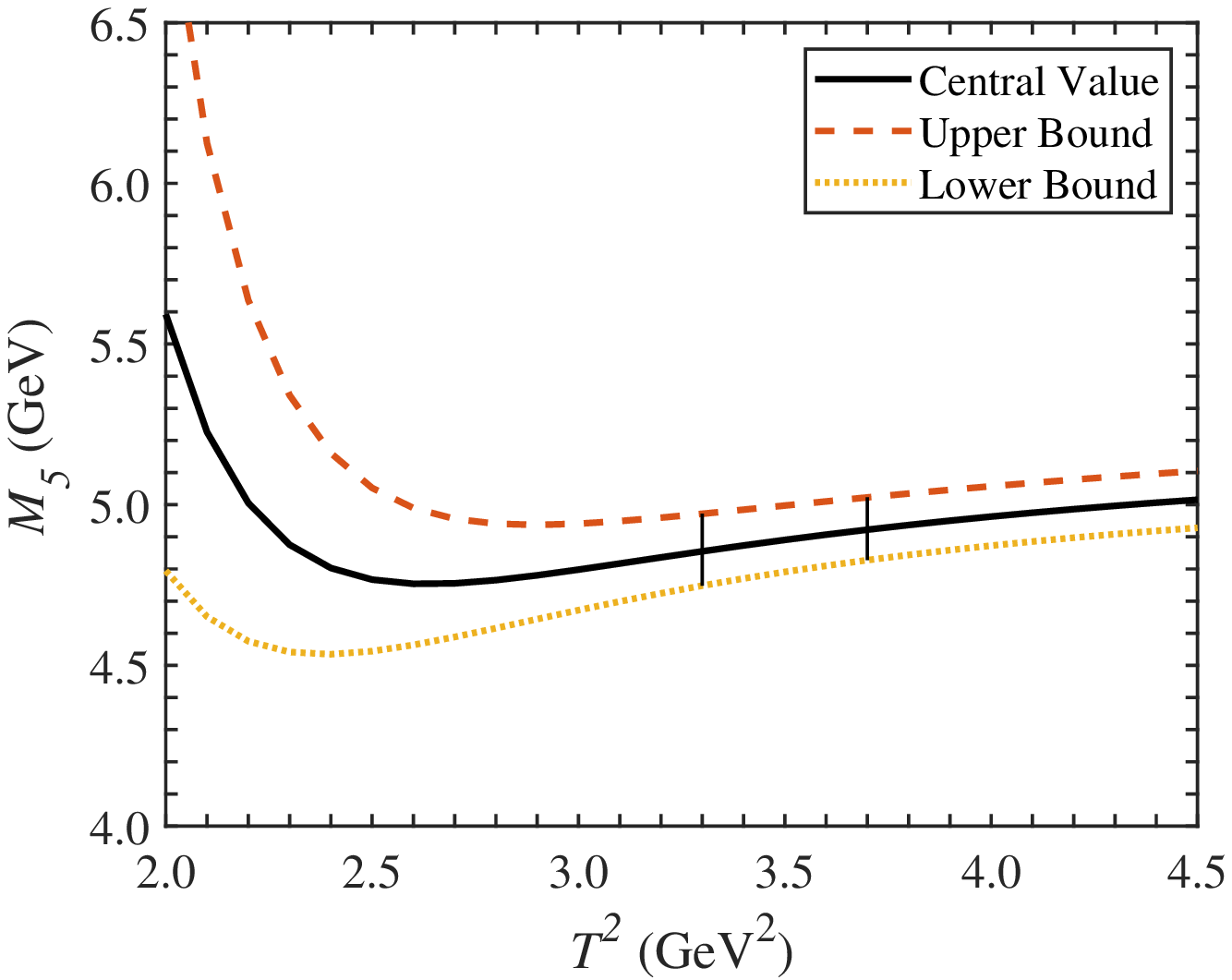}
\end{minipage}
}
\subfigure{
\begin{minipage}[h]{4cm}
\centering
\includegraphics[height=4cm,width=4.5cm]{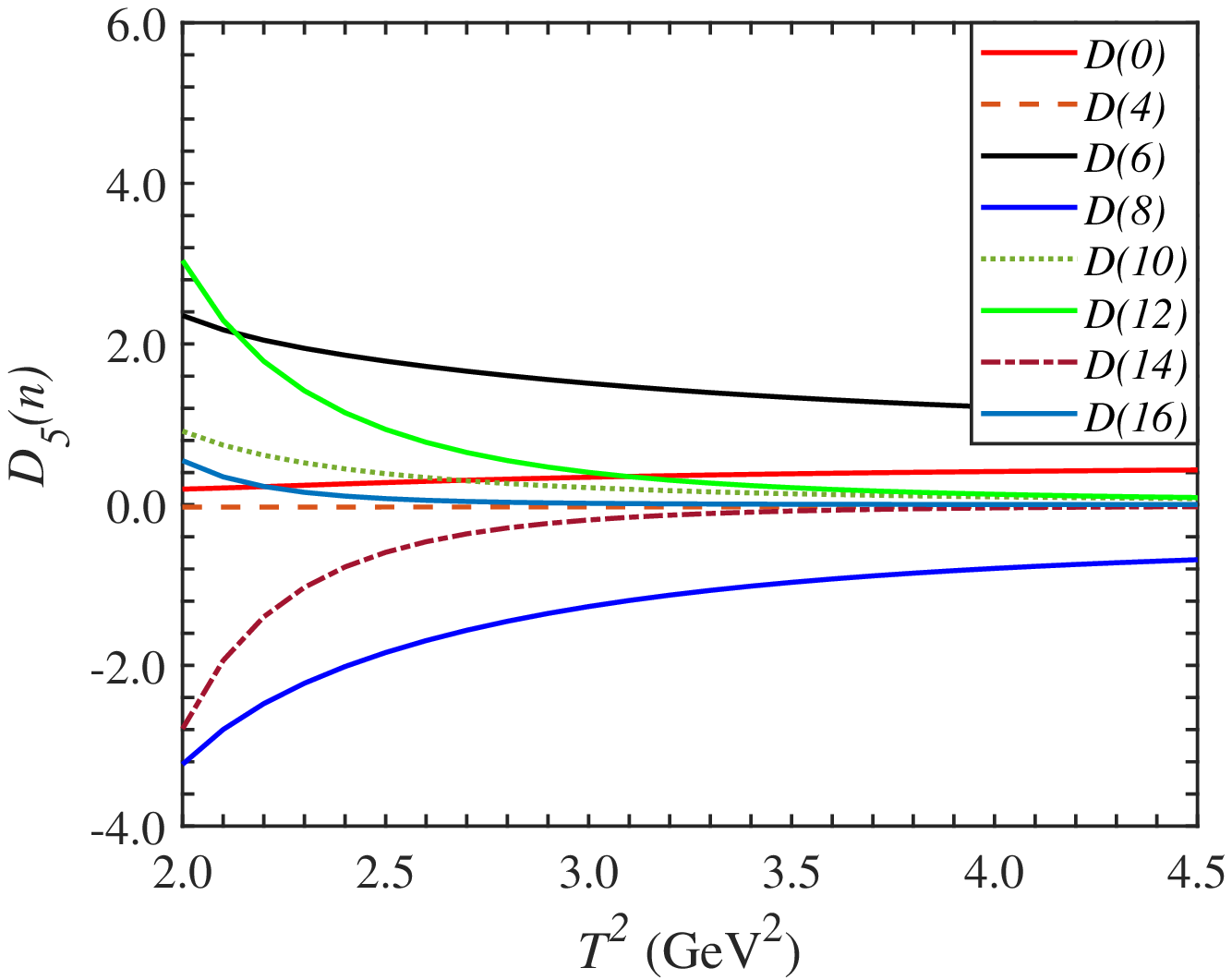}
\end{minipage}
}
\subfigure{
\begin{minipage}[h]{4cm}
\centering
\includegraphics[height=4cm,width=4.5cm]{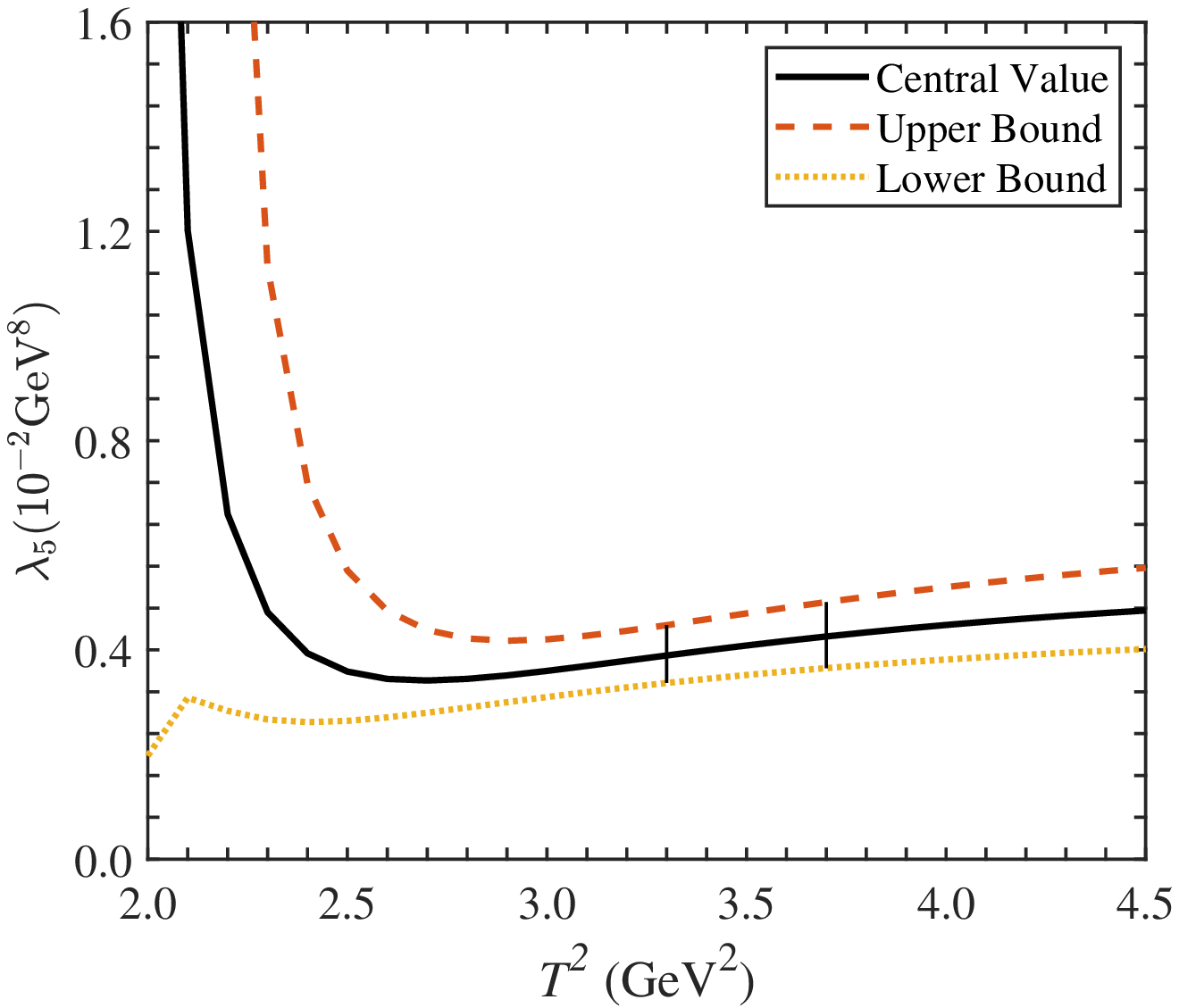}
\end{minipage}
}
\caption{The numerical results of $J_5$: $M_5-T^2$ for the left graph, dashed curve is drawn from the upper values of the input parameters and dotted curve is from the lower values; $D_5(n)-T^2$ for the middle graph; $\lambda_5-T^2$ for the right graph, dashed curve is drawn from the upper values of the input parameters and dotted curve is from the lower values. \label{your label}}
\end{figure}

\begin{figure}[htpb]
\centering
\subfigure{
\begin{minipage}[h]{4cm}
\centering
\includegraphics[height=4cm,width=4.5cm]{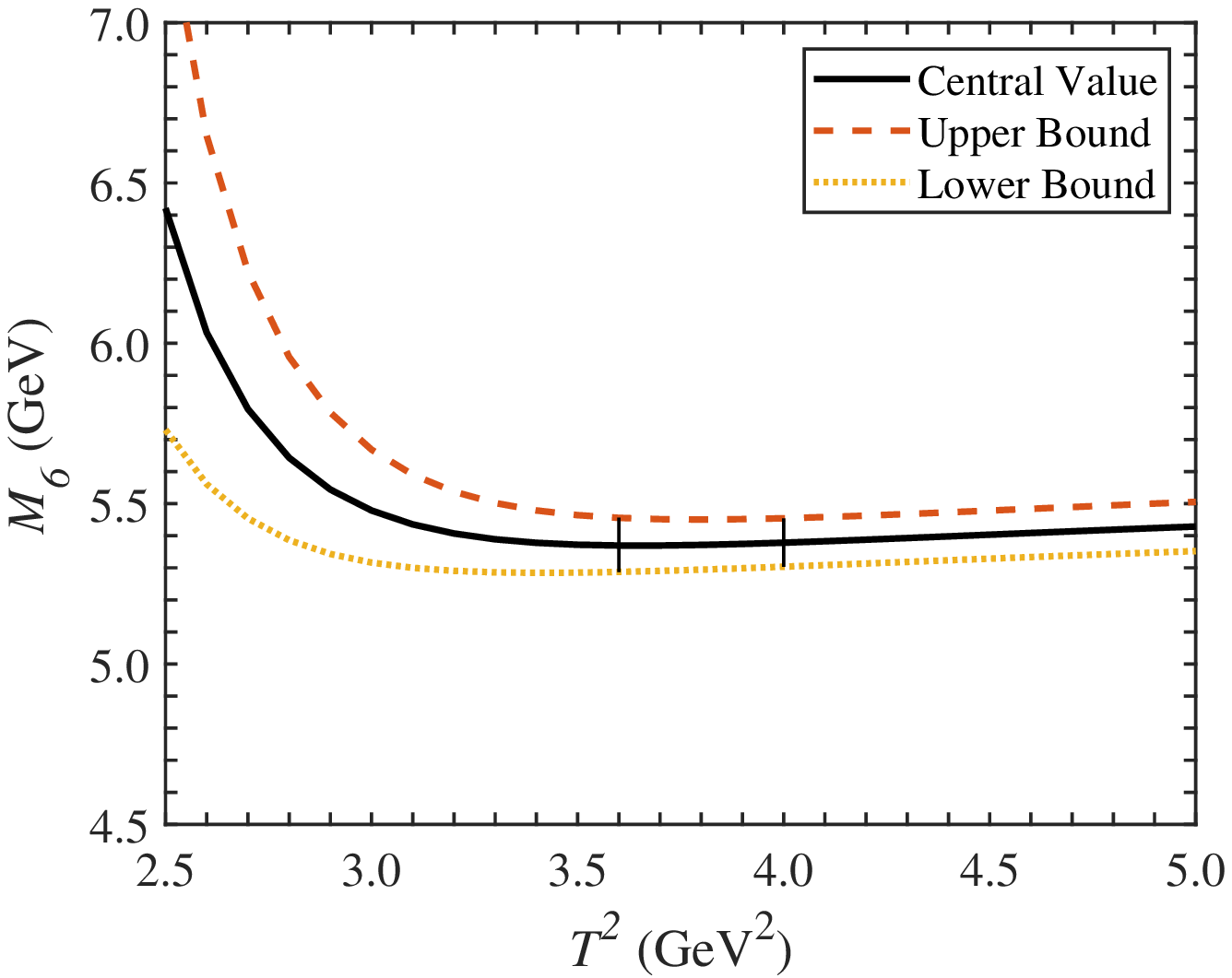}
\end{minipage}
}
\subfigure{
\begin{minipage}[h]{4cm}
\centering
\includegraphics[height=4cm,width=4.5cm]{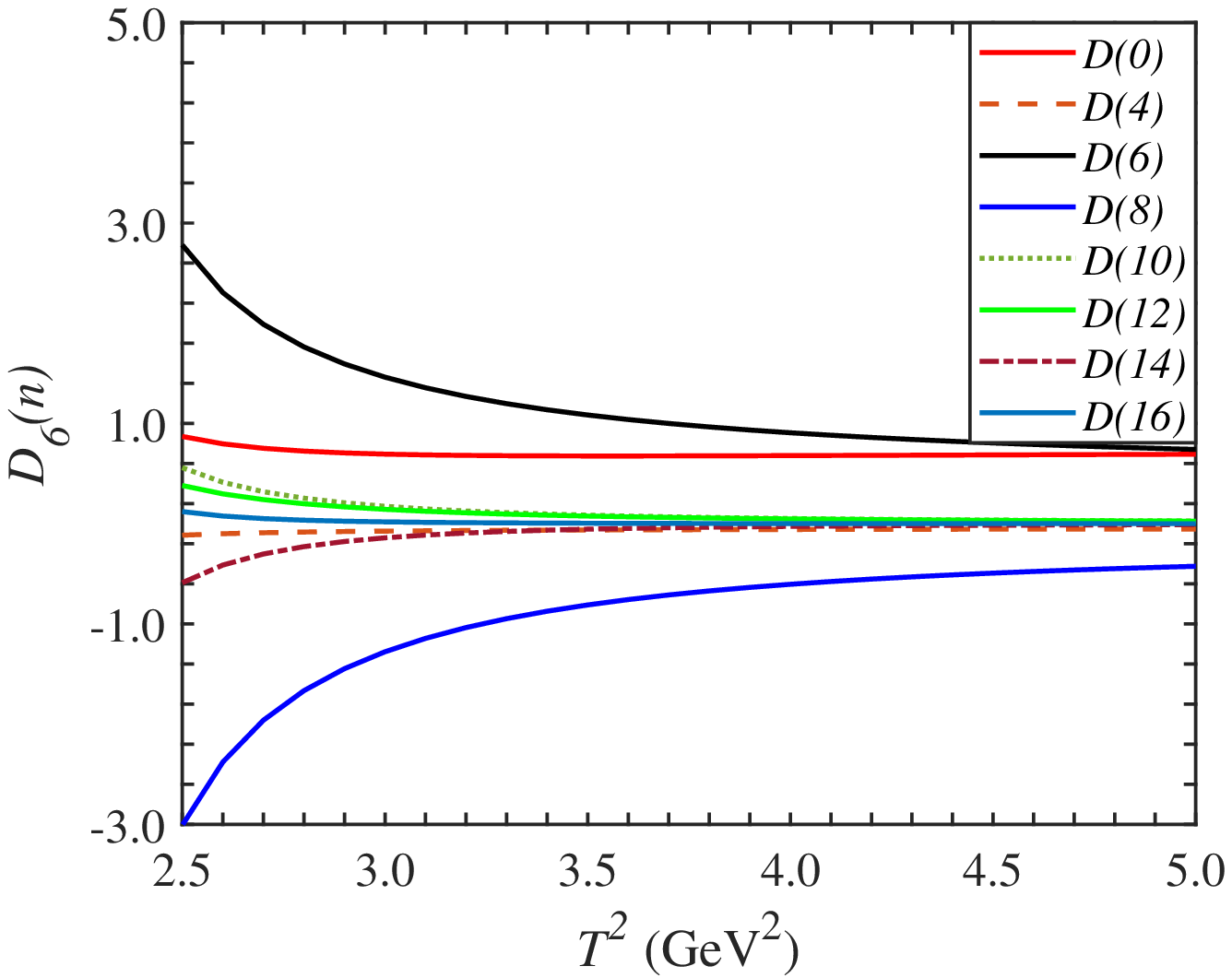}
\end{minipage}
}
\subfigure{
\begin{minipage}[h]{4cm}
\centering
\includegraphics[height=4cm,width=4.5cm]{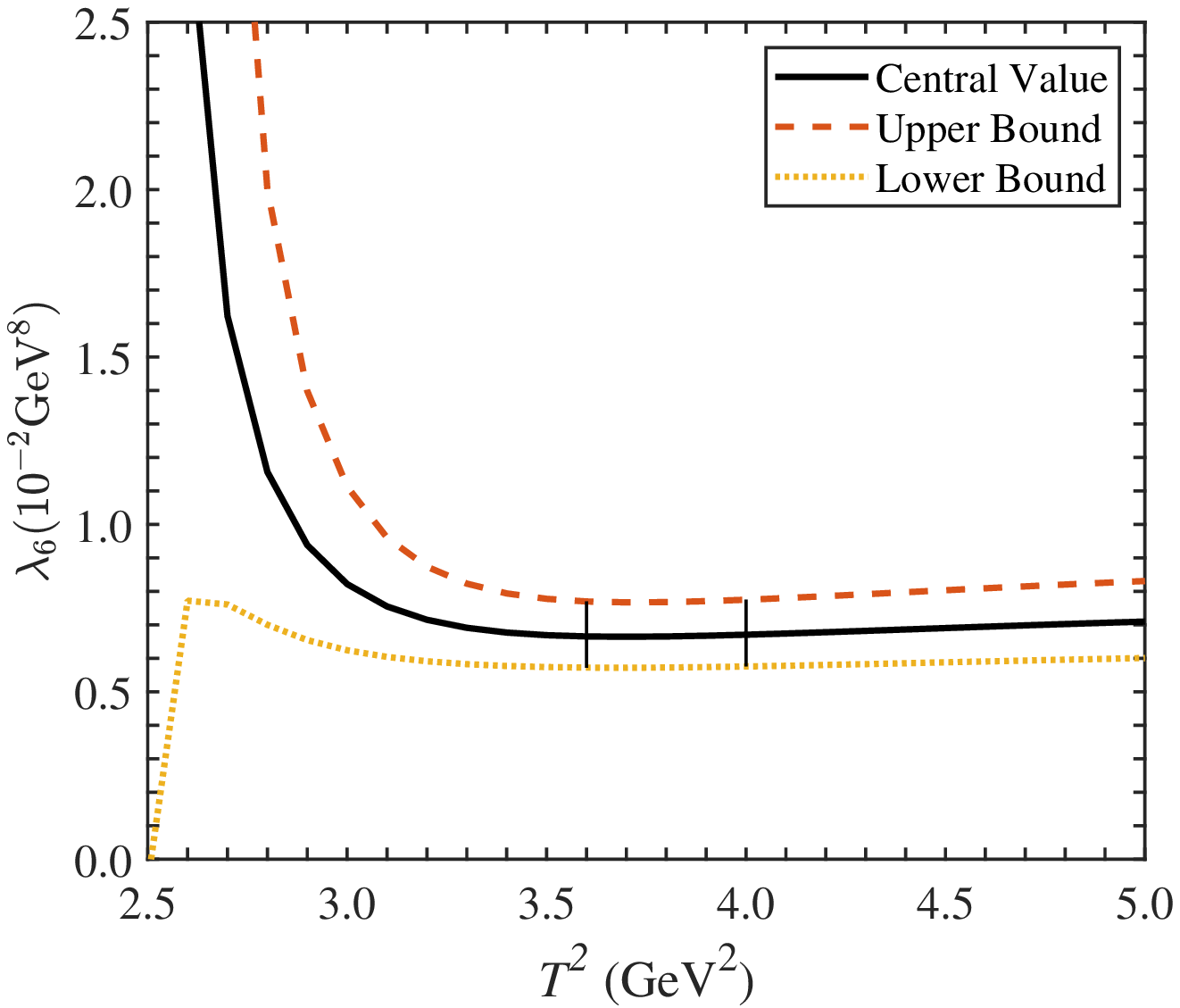}
\end{minipage}
}
\caption{The numerical results of $J_6$: $M_6-T^2$ for the left graph, dashed curve is drawn from the upper values of the input parameters and dotted curve is from the lower values; $D_6(n)-T^2$ for the middle graph; $\lambda_6-T^2$ for the right graph, dashed curve is drawn from the upper values of the input parameters and dotted curve is from the lower values. \label{your label}}
\end{figure}

\begin{figure}[htpb]
\centering
\subfigure{
\begin{minipage}[h]{4cm}
\centering
\includegraphics[height=4cm,width=4.5cm]{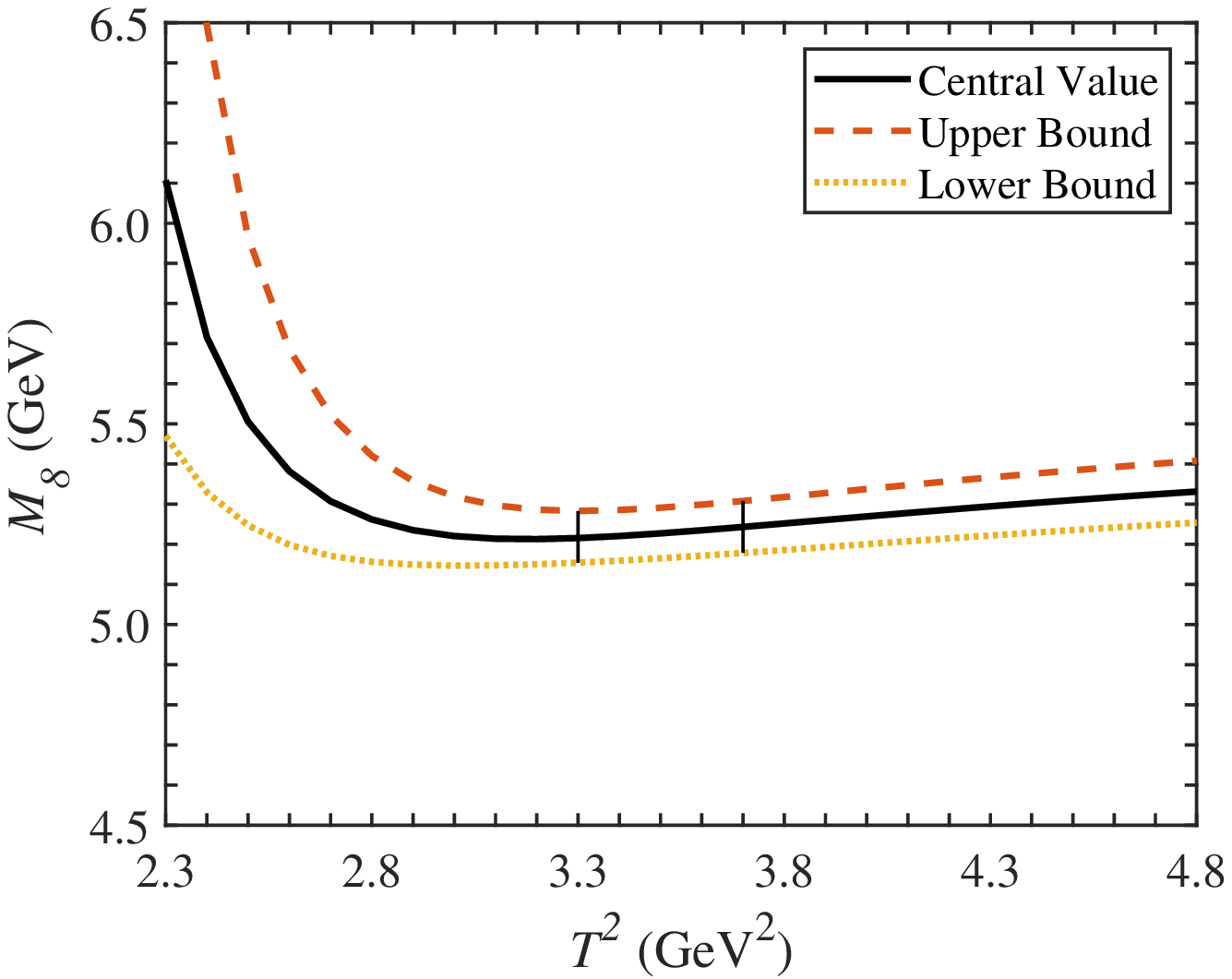}
\end{minipage}
}
\subfigure{
\begin{minipage}[h]{4cm}
\centering
\includegraphics[height=4cm,width=4.5cm]{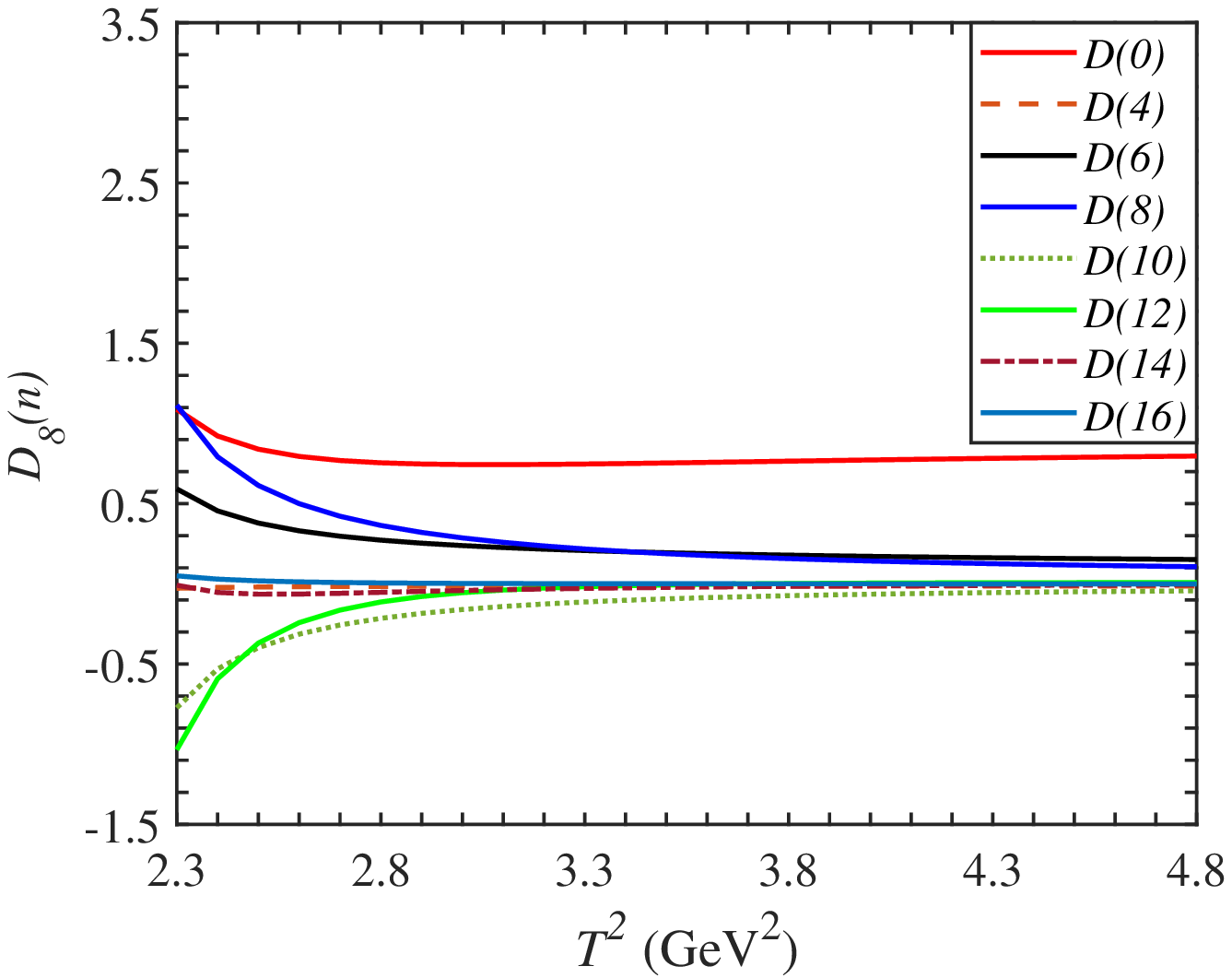}
\end{minipage}
}
\subfigure{
\begin{minipage}[h]{4cm}
\centering
\includegraphics[height=4cm,width=4.5cm]{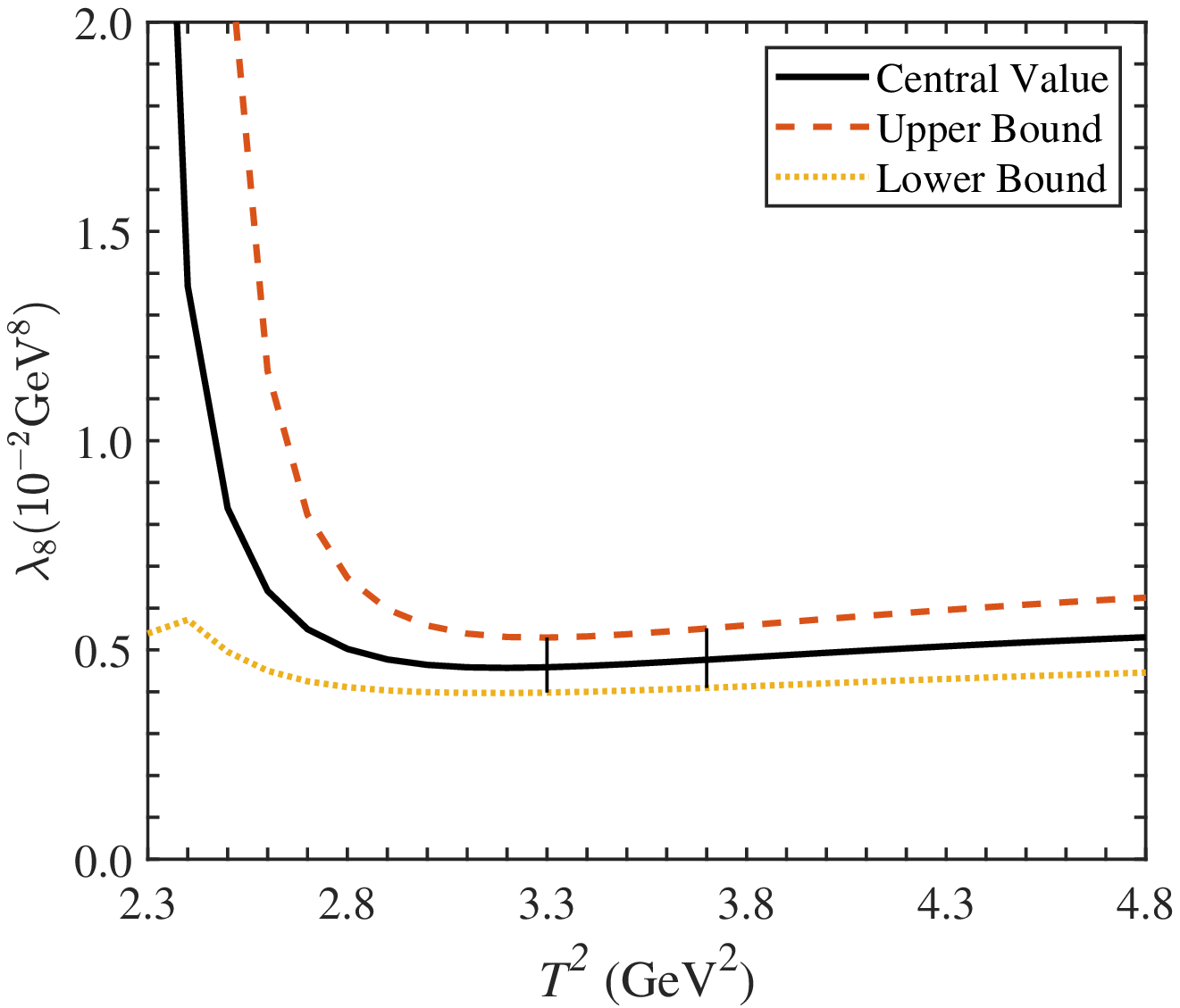}
\end{minipage}
}
\caption{The numerical results of $J_8$: $M_8-T^2$ for the left graph, dashed curve is drawn from the upper values of the input parameters and dotted curve is from the lower values; $D_8(n)-T^2$ for the middle graph; $\lambda_8-T^2$ for the right graph, dashed curve is drawn from the upper values of the input parameters and dotted curve is from the lower values. \label{your label}}
\end{figure}

Due to the uncertainties of the input parameters, the uncertainties of the extracted masses and pole resides are determined by the formula,
\begin{eqnarray}
\delta=\sqrt{\sum\limits_i(\frac{\partial f}{\partial x_i})^2|_{x_i=\bar x_i}(x_i-\bar x_i)^2}\, ,
\end{eqnarray}
where $f$ represent the masses or the pole resides, $x_i$ are the input parameters $\langle\bar qq\rangle$, $\langle\overline{q}g_s\sigma Gq\rangle$, $\cdot\cdot\cdot$, $\bar x_i$ denote the central values of these parameters. Since it is difficult to find the analytic expressions of the partial derivatives $\frac{\partial f}{\partial x_i}$, we carry out the approximate numerical calculations for the uncertainties via,
\begin{eqnarray}
\notag\left(\frac{\partial f}{\partial x_i}\right)^2(x_i-\bar x_i)^2\approx \left[f(\bar x_i\pm \Delta x_i)-f(\bar x_i)\right]^2\, .
\end{eqnarray}

 Except $J_7$, we find the Borel windows for the other seven currents. Their numerical results are shown in Fig.(1)-Fig.(7), respectively. In each figure, we show the masses, dimensional contributions of vacuum condensates and pole residues of each current on $M-T^2$, $D(n)-T^2$ and $\lambda-T^2$ graphs, respectively. Currents $J_1$, $J_2$, $J_3$ and $J_4$ are related to the $\Sigma_c$$\bar{\Sigma}_c$ baryonium structures. For the vector current $J_1$, the extracted mass of the corresponding $\Sigma_c$$\bar{\Sigma}_c$ state with the $J^P=1^-$ is 4.88 GeV with the upper uncertainty being 0.09 GeV. Since the threshold of the two $\Sigma_c$ and $\bar{\Sigma}_c$ baryon constituents is 4.91 GeV, so, we could not say for sure whether it is a molecular state formed by the two baryons or not. On the safe side, it is a possible molecular or compact six-quark state. In the same way, $\Sigma_c$$\bar{\Sigma}_c$ with the $J^P=0^-$ is also a possible molecular or compact six-quark state. As for axial-vector current $J_2$ and scalar current $J_3$, their extracted masses of the corresponding states with the $J^P=1^+$ and $0^+$ are 5.31 GeV and 5.23 GeV, respectively. They are all above the thresholds of the two $\Sigma_c$ and $\bar{\Sigma}_c$ baryon constituents even if we take into the account of the lower bounds of uncertainties of the extracted masses. Both these two are the baryonium resonance states.

Currents $J_5$, $J_6$, $J_7$ and $J_8$ are related to the $\Sigma_c$$\Sigma_c$ dibaryon states. For the current $J_{7,\alpha}=J_{\Sigma}^TC\gamma_{\alpha}J_{\Sigma}$, its related correlation function $\Pi_{7;\alpha\beta}(p)\equiv0$. It indicates that there is no such $\Sigma_c\Sigma_c$ dibaryon state with the $J^P=1^+$. The extracted mass from $J_5$ is $4.89$ {\rm GeV} which is $20$ {\rm MeV} below the the threshold of the two $\Sigma_c$ baryon constituents. Due to the uncertainties of input parameters, the related uncertainty of the mass of this state is up to $0.1$ GeV, it is also a possible $\Sigma_c\Sigma_c$ molecular or a compact six-quark state with the $J^P=0^+$. We also get the masses of $J_6$ and $J_8$, they are all above the thresholds of the two baryon constituents. The detailed results of each states are listed in the Table \uppercase\expandafter{\romannumeral1}.

 For all these seven currents, it is clear to be seen from the $D_i(n)-T^2$ curves in each titled figures, the most important items are the leading order, $\langle\bar qq\rangle^2$ and $\langle\bar qg_s\sigma Gq\rangle\langle\bar qq\rangle$. We find that the higher dimensional vacuum condensates $(n\geq 10)$ play the less important roles. So, the convergence of the operator product expansion holds very well. It is worth mentioning that condensate $\langle\bar qq\rangle^2$ plays a dominant role for the current $J_4$, what's more, the leading order of $J_8$ contributes the most important part for both the mass and pole residue of the corresponding state.

\begin{table*}[t]
\begin{ruledtabular}\caption{The related numerical results extracted from the Borel windows of the seven currents. }
\begin{tabular}{c c c c c c c c c}
& \ $J_i$ & \ $J^P$ &\ $T^2 ({\rm GeV}^2)$    &\ $\sqrt{s_0}({\rm GeV})$ &\ $\mu ({\rm GeV})$ &\ $M({\rm GeV})$ &\ $\rm PC$ &\  $\lambda(10^{-3}{\rm GeV}^8)$ \\
\hline
$\Sigma_c$$\bar{\Sigma}_c$ & \ $J_1$ & \ $1^-$ & \ $3.6\sim4.0$  &  \   $5.60\pm0.1$  & \ $3.2$  & \  $4.88^{+0.09}_{-0.08}$ & \ $(54\sim41)\% $  &  \   $3.23^{+0.46}_{-0.43}$ \\
$\Sigma_c$$\bar{\Sigma}_c$ & \ $J_2$ & \ $1^+$ & \ $3.6\sim4.0$  &  \   $5.90\pm0.1$  & \ $3.8$  & \  $5.31^{+0.07}_{-0.07}$ & \ $(54\sim42)\% $  &  \   $4.74^{+0.72}_{-0.64}$ \\
$\Sigma_c$$\bar{\Sigma}_c$ & \ $J_3$ & \ $0^+$ & \ $3.4\sim3.8$  &  \   $5.80\pm0.1$  & \ $3.7$  & \  $5.23^{+0.07}_{-0.07}$ & \ $(55\sim43)\% $  &  \   $4.19^{+0.67}_{-0.59}$ \\
$\Sigma_c$$\bar{\Sigma}_c$ & \ $J_4$& \ $0^-$ & \ $3.6\sim4.0$  &  \   $5.60\pm0.1$  & \ $3.2$  & \  $4.88^{+0.08}_{-0.08}$ & \ $(53\sim40)\% $  &  \   $3.34^{+0.48}_{-0.45}$ \\
\hline
$\Sigma_c$$\Sigma_c$ & \ $J_5$ & \ $0^+$ & \ $3.3\sim3.7$  &  \   $5.55\pm0.1$  & \ $3.2$  & \  $4.89^{+0.11}_{-0.10}$ & \ $(56\sim42)\% $  &  \   $4.09^{+0.61}_{-0.56}$ \\
$\Sigma_c$$\Sigma_c$ & \ $J_6$& \ $0^-$ & \ $3.6\sim4.0$  &  \   $5.90\pm0.1$  & \ $3.9$  & \  $5.37^{+0.08}_{-0.08}$ & \ $(52\sim40)\% $  &  \   $6.65^{+1.03}_{-0.93}$ \\
$\Sigma_c$$\Sigma_c$ & \ $J_8$& \ $1^-$ & \ $3.3\sim3.7$  &  \   $5.85\pm0.1$  & \ $3.7$  & \  $5.23^{+0.06}_{-0.06}$ & \ $(57\sim43)\% $  &  \   $4.66^{+0.71}_{-0.63}$ \\
\hline
$\Lambda_c$$\Lambda_c$ & \ $J_1$ & \ $0^+$ & \ $3.4\sim3.8$  &  \   $5.70\pm0.1$  & \ $3.6$  & \  $5.12^{+0.15}_{-0.12}$ & \ $(54\sim42)\% $  &  \   $1.81^{+0.31}_{-0.27}$ \\
$\Lambda_c$$\bar{\Lambda}_c$ & \ $J_2$ & \ $0^-$ & \ $3.2\sim3.6$  &  \   $5.30\pm0.1$  & \ $2.8$  & \  $4.66^{+0.10}_{-0.09}$ & \ $(54\sim41)\% $  &  \   $0.64^{+0.10}_{-0.09}$ \\
$\Lambda_c$$\bar{\Lambda}_c$ & \ $J_3$ & \ $1^+$ & \ $3.4\sim3.8$  &  \   $5.60\pm0.1$  & \ $3.4$  & \  $5.00^{+0.10}_{-0.09}$ & \ $(55\sim42)\% $  &  \   $0.64^{+0.10}_{-0.09}$ \\
$\Lambda_c$$\bar{\Lambda}_c$ & \ $J_4$ & \ $1^-$ & \ $3.2\sim3.6$  &  \   $5.25\pm0.1$  & \ $2.7$  & \  $4.57^{+0.09}_{-0.09}$ & \ $(53\sim39)\% $  &  \   $0.52^{+0.09}_{-0.08}$ \\
\end{tabular}
\end{ruledtabular}
\end{table*}

In 2017, the LHCb has observed the doubly-charmed baryon $\Xi_{cc}^{++}$ with mass of $3621.40\pm0.78\rm MeV$  \cite{Raaij}. It is proposed that there is large binding energy $B(cc)=129\rm MeV$ of the two heavy c-quarks in a baryon \cite{Marek}. In Ref.  \cite{Marek2}, the $quark-level$ analogue of nuclear fusion for the two $\Lambda_c$ baryons was discussed. In a similar way, we also consider the quark-rearrangement reaction, the possible decays of $\Sigma_c\Sigma_c$ dibaryon states are given by,
\begin{eqnarray}
\notag  \Sigma_c\Sigma_c &&\rightarrow  \Xi_{cc}^{++}n\, ,\\
 cud\quad cud && \rightarrow ccu\quad ddu \, .\\
\notag
\end{eqnarray}
Since the $J^P$ of $\Xi_{cc}^{++}$ is unknown yet, thus, it is also impossible for us to determine the $J^P$ of the produced neutron after the decay. We can only estimate the upper values of the exothermic energy $Q_{upper}$ of each $\Sigma_c\Sigma_c$ quark-rearrangement reaction by setting the minimum mass of the neutron as $m_n=1440\rm MeV$. For $\Sigma_c\Sigma_c$ with the $J^P=0^+$, it is impossible to have such kind of decay since its mass is less than the sum of $m_{\Xi_{cc}^{++}}$ and $m_n$. As for $0^-$ and $1^-$, the corresponding $Q_{upper}$ are 310 MeV and 170 MeV, respectively.

\begin{large}
\noindent\textbf{4 Conclusions}
\end{large}

In this article, we construct the eight interpolating currents to study the $\Sigma_c\bar{\Sigma}_c$ baryonium and $\Sigma_c\Sigma_c$ dibaryon states via the QCD sum rules. The correlation function of $J_7$ is zero which indicates that there is no such $\Sigma_c$$\Sigma_c$ dibaryon state with the $J^P=1^+$. We find two possible $\Sigma_c\bar{\Sigma}_c$ molecular or compact six-quark states with the $J^P=1^-$ and $0^-$. Their masses are $4.88^{+0.09}_{-0.08}$ GeV and $4.88^{+0.08}_{-0.08}$ GeV, respectively. For the currents $J_2$, $J_3$, $J_6$ and $J_8$, the extracted masses of these four states are $5.31^{+0.07}_{-0.07}$ GeV, $5.23^{+0.07}_{-0.07}$ GeV, $5.37^{+0.08}_{-0.08}$ GeV and $5.23^{+0.06}_{-0.06}$ GeV, respectively. Both the four are above the the thresholds of the two baryon constituents. We conclude that $\Sigma_c\bar{\Sigma}_c$ with the $J^P=1^+$ and $0^+$ are baryonium resonances states and $\Sigma_c\Sigma_c$ with the $J^P=0^-$ and $1^-$ are dibaryon resonances states. We find the central value of the mass of the state with the $J^P=0^+$ is just $20$ MeV below the threshold of the two $\Sigma_c$ baryon constituents. Considering the uncertainty, $\Sigma_c\Sigma_c$ dibaryon with the $J^P=0^+$ is also a possible molecular state. For all the bound states related to $\Sigma_c\bar{\Sigma}_c$ and $\Sigma_c\Sigma_c$, we need more data, such as the electromagnetic form factors, to differentiate whether the bound dibaryon or baryonium is a molecular state or a compact six-quark entity. This is also the task of our further work. The possible decays, the quark-rearrangement reactions of each $\Sigma_c\Sigma_c$ states are discussed. The upper values of the exothermic energy are calculated for these decays which may be experimentally verified in the future.

\begin{large}
\noindent\textbf{Data Availability}
\end{large}

\noindent All data included in this manuscript are available upon request by contacting with the correspond-
ing author.

\begin{large}
\noindent\textbf{Conflicts of Interest}
\end{large}

\noindent The authors declare that they have no conflicts of interest.


\begin{large}
\noindent\textbf{Acknowledgments}
\end{large}

\noindent This work has been supported by National Natural Science Foundation, Grant Number 12175068 and Youth Foundation of NCEPU, Grant Number 93209703.

\end{document}